# Pressure-Driven Evolution of the Physical Properties of Layered $Y_2GaB$ and $Y_2InB$ MAB Phases: A Comparative First-Principles Study

M. Rabi Us Sani Shapon, Md. Asif Afzal, Jubair Hossan Abir, Raihana Shams Islam*, Saleh Hasan Naqib*

Department of Physics, University of Rajshahi, Rajshahi 6205, Bangladesh

*Corresponding authors; Emails: jsr_eda1@ru.ac.bd (Saleh Hasan Naqib), rsislam@ru.ac.bd (Raihana Shams Islam)

## Abstract

MAB phases are nanolaminated ternary borides in which strong in-plane M-B bonds and weaker interlayer interactions drive their distinctive stability and functionality. Hydrostatic pressure offers a direct route to tune the in-plane and interlayer interactions that govern the stability and functionality of layered $Y_2GaB$ and $Y_2InB$ borides. Here, first-principles calculations are used to compare their structural, elastic, thermophysical, electronic, bonding, optical, and vibrational responses from 0 to 20 GPa. Elastic criteria show that $Y_2GaB$ is stable at 0, 5, and 10 GPa, whereas $Y_2InB$ is stable only at ambient pressure. Within the stable $Y_2GaB$ range, compression increases the elastic moduli, acoustic velocities, Debye temperature, and lattice thermal conductivity while reducing the elastic anisotropy. Both compounds are multiband metals and spin-orbit coupling modifies local dispersions without opening a gap. $Y_2GaB$ retains nine Fermi-surface sheets from 0 to 10 GPa, whereas $Y_2InB$ has eight at ambient pressure, with no pressure-induced topological reconstruction. Charge-density and overlap analyses indicate mixed ionic-covalent Y-B bonding within a delocalized metallic background. Optical parameters are anisotropic. Unit-cell phonon spectra contain positive frequencies at 0 and 10 GPa, although expanded-supercell calculations reveal low-frequency sensitivity. These results establish mechanically viable pressure windows and clarify substitution-driven structure-property relationships for these borides.

## 1 Introduction

MAB phases are layered ternary borides built from metal-boron frameworks interleaved with A-element containing layers. Their crystal chemistry is not confined to a single prototype: composition, stacking sequence, symmetry, boron coordination, and the connectivity of the metal-boron sublattice vary substantially across the families. Strong directional metal-boron interactions commonly coexist with weaker or more delocalized bonding associated with the A-containing layers which produces metallic electronic behavior together with ceramic-like stiffness and direction-dependent mechanical response. This nanolaminated character is closely related to the broader MAX-phase framework, while retaining a chemically distinct bonding and compositional identity **[1–5]**. Structural optimization alone is insufficient to establish phase stability; a comprehensive evaluation of energetic, elastic, and vibrational characteristics is therefore essential for candidate layered compounds [6]. Moreover, theoretical and experimental studies of boron-containing layered ternaries indicate that substituting carbon or nitrogen with boron is not merely an isostructural replacement, as it can induce distinct coordination environments and fundamentally different bonding networks [7,8].

The scientific importance of MAB compounds lies in this combination of structural diversity and chemically tunable bonding. Transition-metal selection, A-site chemistry, layer stacking, symmetry, and external pressure can alter the relative strengths of in-plane and interlayer interactions, thereby changing elastic stiffness, anisotropy, electronic states, charge redistribution, optical response, and lattice dynamics. Recent developments encompass topochemical access to two-dimensional MBene layers [9], experimentally synthesized $Nb_2SB$, $Zr_2SB$, and $Hf_2SB$ phases [10,11], molten-salt production of MoB MBene nanosheets [12], and broader theoretical and experimental studies of related layered borides [13–15]. Together, these advances illustrate the structural and chemical breadth of the MAB family. These systems therefore provide a useful platform for examining how local coordination and layer chemistry control the coexistence of metallic and ceramic characteristics in anisotropic solids, rather than supporting a single universal description of MAB behavior.

Suitable layered MAB compounds can serve as precursors to two-dimensional boride derivatives through selective removal of suitable A-containing layers. Experimental deintercalation of MoAlB and related intermediate structures has demonstrated access to MoB-based MBene layers and nanosheets through both topochemical and molten-salt routes [9,12]. Such derivatives should not be assumed to reproduce the properties of either the parent bulk phase or conventional carbide/nitride MXenes - their feasibility and response depend on precursor stability, cleavage or exfoliation pathways, surface termination, composition, and the bonding that links the removable layer to the boride framework. Not every MAB compound is therefore expected to be exfoliable, and no experimentally established MABene derivative of $Y_2GaB$ or $Y_2InB$ is assumed here. Understanding the stability, bonding and pressure response of the parent phases remains a necessary foundation for any future assessment of reduced dimensional derivatives.

$Y_2GaB$ and $Y_2InB$ were previously predicted in the layered hexagonal $P6_3/mmc$ structure, where Y-B-based layers are separated or modulated by Ga- or In-containing planes [16]. Their common framework enables a controlled comparison of Ga-to-In substitution while retaining the same nominal structural family. Replacing Ga with In changes the lattice dimensions, atomic-mass distribution, A-site *p*-state energies, and local interactions without altering the basic stacking motif. The previous compound-specific study reported their ambient-pressure structural, elastic, electronic, bonding, optical, and phonon characteristics; in particular, it reported negative-frequency phonon modes under its adopted computational setup [16]. That work established the initial physical picture but did not provide a unified pressure-dependent comparison or resolve how alternative phonon representations affect the stability assessment.

Several compound-specific questions therefore remain open. It is necessary to determine how hydrostatic compression changes the basal and axial lattice dimensions, whether the two phases retain the same elastic-stability window under pressure-corrected criteria, and how stiffness, ductility indicators, and directional anisotropy evolve before instability occurs. Related layered compounds show that pressure can modify elastic and electronic behavior non-uniformly [17–19]. Aggregate elastic anisotropy can be quantified using the universal anisotropy index [20], although direction-resolved analysis is required to identify the underlying directional response. The associated changes in acoustic velocities, elastic Debye temperature, empirical melting-temperature estimates, minimum thermal-conductivity measures, and Slack-based lattice transport also require comparison within the elastically stable states. At the electronic level, unresolved issues include SOC-sensitive band rearrangements, redistribution of the near-Fermi level electronic energy density of states (DOS), and whether pressure creates, removes, or reconnects Fermi-surface sheets. The evolution of charge density, thermal parameters, Y-B bonding, and polarization-dependent optical spectra under pressure and Ga-to-In substitution is likewise unknown. Finally, elastic stability against homogeneous strain [21] and phonon stability against wave-vector-dependent displacements [22] must be assessed separately, including the sensitivity of the phonon conclusion to unit-cell and expanded-supercell treatments.

Motivated by these gaps, the present work evaluates pressure and Ga-to-In substitution within a consistent first-principles framework. Structural evolution is examined from 0 to 20 GPa, while pressure-corrected elastic stability, mechanical properties, and elastic anisotropy are determined across the relevant states. The pressure range has been selected considering its accessibility in general pressure-study laboratories. Elastic-derived thermophysical quantities-including sound velocities, Debye temperature, an empirical melting-temperature estimate, Clarke minimum thermal conductivity, and Slack-estimated lattice thermal conductivity-are evaluated only for $Y_2GaB$ at 0, 5, and 10 GPa and $Y_2InB$ at 0 GPa, considering the dynamic stability of these two compounds. Most of these quantities retain the static-lattice basis of the underlying DFT calculations; only the Slack lattice contribution is estimated at 300 K for comparative purposes, and the metallic character implies that an electronic contribution may also enter the total thermal conductivity. The electronic analysis combines band structures with and without SOC, total and partial densities of states, and Fermi-surface topology. Bonding is examined through electron-density differences and selected bond-overlap evidence, and the optical response is resolved along [100] and [001] directions of electric field polarization. Unit-cell phonon dispersions are calculated at 0 and 10 GPa for both compounds, whereas expanded-supercell calculations are used only to assess sensitivity to cell size and force-constant treatment.

## 2 Computational Methodology

First Principle calculations were performed within Kohn-Sham density-functional theory (DFT) [23,24] using CASTEP [25] and the Perdew, Burke, Ernzerhof (PBE) generalized-gradient approximation (GGA) [26]. The calculations employed on-the-fly-generated (OTFG) ultrasoft pseudopotentials [27] with scalar-relativistic treatment [28], a 400 eV plane-wave cutoff, a 15 × 15 × 3 Monkhorst-Pack mesh [29], and 0.1 eV Gaussian smearing. The hexagonal $P6_3/mmc$ structures were optimized under hydrostatic pressures ranging from 0 to 20 GPa in 5 GPa steps using BFGS [30] algorithm. The convergence thresholds were $5 \times 10^{-6}$ eV atom$^{-1}$ for energy, 0.01 eV Å$^{-1}$ for force, 0.02 GPa for stress, and $5 \times 10^{-4}$ Å for displacement.

Elastic constants were obtained by the finite stress-strain method [31]. Elastic moduli, anisotropy, acoustic velocities, elastic Debye temperature, empirical melting temperature, and Clarke model minimum thermal conductivity were derived from static-lattice DFT data for the mechanically supported states. The Slack model alone was evaluated at 300 K to estimate lattice thermal conductivity. Electronic analyses included band structures, densities of states, and Fermi surfaces. SOC and non-SOC band structures were computed and compared using Quantum ESPRESSO [32] with fully relativistic PAW-PBE potentials [33]. Bonding was examined primarily through electron-density-difference maps, with selected population and bond-overlap results used as complementary evidence [34–36]. Polarization-resolved optical spectra were calculated from the optimized structures [37,38].

Phonons were obtained using finite-displacement method [39]. An 8 × 8 × 2 *k*-point mesh and compound-specific converged cutoffs were used, and spectra were evaluated for both compounds at 0 and 10 GPa. Property-specific equations are reported in the corresponding Results and Discussion subsections.

# 3 Results and discussion

## 3.1 Structural Properties

The optimized $Y_2GaB$ and $Y_2InB$ structures retain the hexagonal $P6_3/mmc$ (No. 194) 211 arrangement, in which Y-B slabs alternate with Ga- or In-containing layers along the *c* axis (Figure 1). This stacking is consistent with related layered hexagonal 211 borides, including $Nb_2SB$, $Zr_2SB$, and $Hf_2SB$, and with calculated $M_2GaB$ analogues [1,10,11,40,41]. At 0 GPa, both compounds have $a$ = 3.64 Å, whereas replacing Ga with In increases $c$ from 16.19 to 16.96 Å and $V$ from 186.70 to 195.03 Å$^3$. The increase in volume results from larger ionic radius of In compared to Ga. The basal dimension is therefore nearly unchanged, while the In-containing layer expands the axial repeat. Despite its larger cell, $Y_2InB$ is denser (5.16 versus 4.59 g cm$^{-3}$) because In has the greater atomic mass. The ambient-pressure lattice constants agree closely with the earlier report ($a$ = 3.63 Å and $c$ = 16.23 Å for $Y_2GaB$; $a$ = 3.66 Å and $c$ = 16.99 Å for $Y_2InB$), with deviations approximately 0.6% [16].

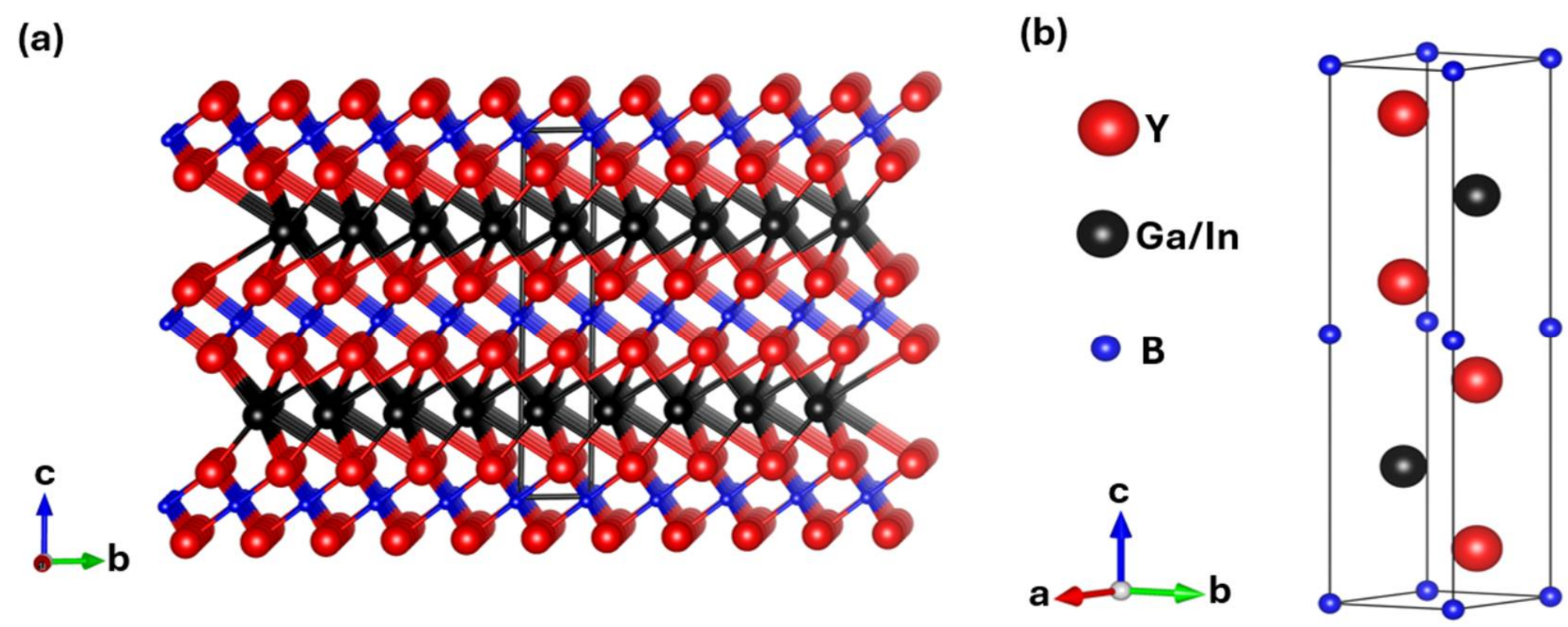

**Figure 1** (a) Layered and (b) three-dimensional representations of the hexagonal *P6₃/mmc* crystal structure of $Y_2AB$, where *A* = Ga or In.

For a hexagonal unit cell, the volume is:

$$V = \frac{\sqrt{3}}{2} a^2 c \tag{1}$$

and the density is:

$$\rho = \frac{ZM}{N_A V \times 10^{-24}} \tag{2}$$

where *Z* is the number of formula units per unit cell (*Z* = 2), *M* is the molar mass, $N_A$ is Avogadro's constant, and *V* is the unit-cell volume in $Å^3$.

**Table 1** Pressure-dependent optimized lattice parameters (*a* = *b* and *c*, in Å), axial ratio *c/a*, unit-cell volume *V* ($Å^3$), density *ρ* (g $cm^{-3}$).

| Phase | Pressure (GPa) | *a* (Å) | *c* (Å) | *c/a* | *V* ($Å^3$) | *ρ* (g $cm^{-3}$) | Reference |
|---|---|---|---|---|---|---|---|
| $Y_2GaB$ | 0 | 3.64 | 16.19 | 4.44 | 186.70 | 4.59 | This work |
| | | 3.63 | 16.23 | 4.47 | 185.34 | - | [16] |
| | 5 | 3.53 | 15.85 | 4.47 | 171.98 | 4.98 | This work |
| | 10 | 3.45 | 15.63 | 4.53 | 161.23 | 5.32 | |
| | 15 | 3.36 | 15.48 | 4.60 | 151.51 | 5.66 | |
| | 20 | 3.30 | 15.40 | 4.65 | 146.03 | 5.87 | |
| $Y_2InB$ | 0 | 3.64 | 16.96 | 4.65 | 195.03 | 5.16 | |
| | | 3.66 | 16.99 | 4.63 | 197.55 | - | [16] |
| | 5 | 3.55 | 16.66 | 4.68 | 182.46 | 5.52 | This work |
| | 10 | 3.46 | 16.52 | 4.77 | 171.52 | 5.87 | |
| | 15 | 3.38 | 16.35 | 4.82 | 162.65 | 6.19 | |
| | 20 | 3.32 | 16.27 | 4.89 | 155.50 | 6.48 | |

The optimized lattice parameters, axial ratios, unit-cell volumes, and densities are listed in **Table 1**. Both lattice parameters decrease smoothly over 0-20 GPa, although the basal plane contracts more strongly. For $Y_2GaB$, *a* decreases from 3.64 to 3.30 Å (approximately 9.3%), whereas $c$ falls from 16.19 to 15.40 Å (approximately 4.9%). $Y_2InB$ follows the same directional pattern: $a$ decreases from 3.64 to 3.32 Å (approximately 8.8%) and $c$ from 16.96 to 16.27 Å (approximately 4.1%). Consequently, $c/a$ increases from 4.44 to 4.65 for $Y_2GaB$ and from 4.65 to 4.89 for $Y_2InB$. The larger $c/a$ of $Y_2InB$ persists at every pressure, while Ga-to-In substitution leaves the qualitative compression pattern unchanged. Comparable direction-dependent contraction has been reported for related layered carbides and borides, although the relative responses of $a$ and $c$ remain dependent on composition and stacking **[17–19,42,43]**.

The unit-cell volume decreases monotonically from 186.70 to 146.03 Å³ for $Y_2GaB$ and from 195.03 to 155.50 Å³ for $Y_2InB$, corresponding to reductions of approximately 21.8% and 20.3%, respectively. Such large decrease in volume is a clear indication of soft nature of the two MAB phases. $Y_2GaB$ undergoes a slightly larger fractional volume contraction over the investigated pressure interval. At the same time, the density rises to 5.87 g cm$^{-3}$ for $Y_2GaB$ and 6.48 g cm$^{-3}$ for $Y_2InB$ and the In-containing compound remains both denser and more voluminous throughout. **Table 1** provides the complete pressure-dependent values, and **Figure 2** summarizes their evolution.

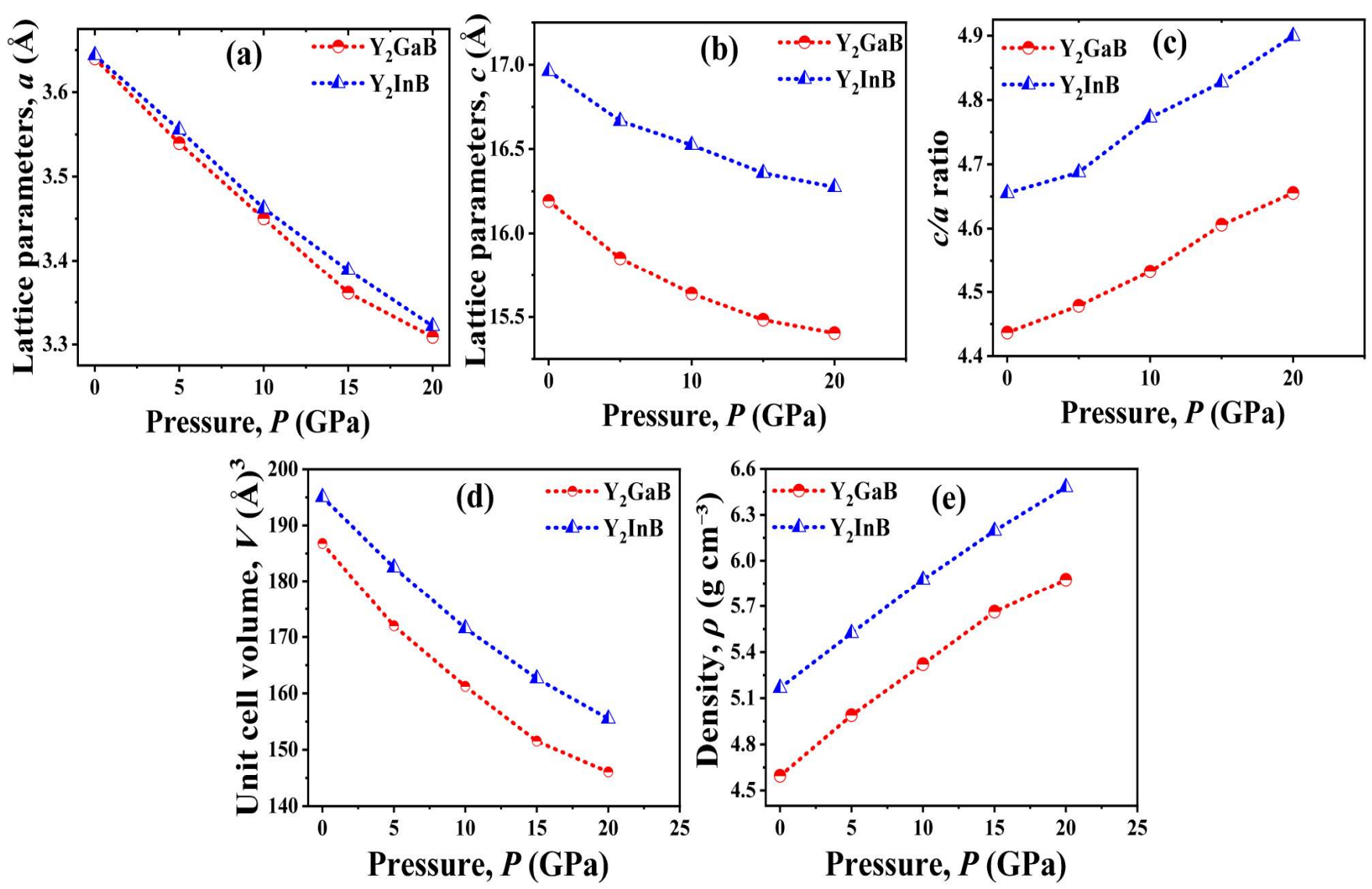

**Figure 2** Pressure dependence of the (a) basal lattice parameter $a$, (b) axial lattice parameter $c$, (c) axial ratio $c/a$, (d) unit-cell volume $V$, and (e) density $\rho$ of $Y_2GaB$ and $Y_2InB$ over 0-20 GPa.

No abrupt discontinuity appears in $a$, $c$, $c/a$, $V$, or $\rho$ at the sampled pressures, indicating continuous compression within the optimized $P6_3/mmc$ structural model. All optimized structures are therefore retained in **Table 1** as structural data, while the elastic-stability windows are determined independently from the pressure-corrected criteria in Section 3.2 and the phonon response is assessed separately in Section 3.10 **[6]**.

## 3.2 Elastic and Mechanical Properties

For hexagonal $Y_2GaB$ and $Y_2InB$, the five independent stiffness coefficients are $C_{11}$, $C_{12}$, $C_{13}$, $C_{33}$, and $C_{44}$; the basal-plane shear coefficient $C_{66}$ is obtained from **Eq. (3) [21]**. The calculated single-crystal elastic constants and stability classifications are summarized in **Table 2**.

$$C_{66} = \frac{(C_{11} - C_{12})}{2} \tag{3}$$

Elastic stability under hydrostatic pressure was assessed from the stress-corrected stiffness tensor **[44]**. For hexagonal symmetry, the required finite-pressure conditions are **[21,44]**:

$$C_{44} - P > 0 \tag{4}$$

$$C_{66} - P > 0 \tag{5}$$

$$(C_{11} + C_{12})(C_{33} - P) - 2(C_{13} + P)^2 > 0 \tag{6}$$

where $P$ is the applied hydrostatic pressure. $Y_2GaB$ satisfies **Eqs. (4)-(6)** at 0, 5, and 10 GPa but fails the adopted criteria at 15 and 20 GPa. $Y_2InB$ satisfies all three conditions only at 0 GPa. **Table 2** represents the elastically stable states, and the complete unstable-state tensors are not reproduced here. Subsequent moduli and mechanical descriptors are evaluated only for $Y_2GaB$ at 0, 5, and 10 GPa and $Y_2InB$ at 0 GPa. This classification concerns stability against homogeneous strain and is independent of the phonon assessment in Section 3.10.

**Table 2** Calculated elastic constants $C_{ij}$, basal-plane shear constant $C_{66} = (C_{11} - C_{12})/2$, and Cauchy pressures $C^{P}_{1} = C_{13} - C_{44}$ and $C^{P}_{2} = C_{12} - C_{66}$ for the elastically stable states of $Y_2GaB$ and $Y_2InB$. All elastic quantities are in GPa.

| Compound | $P$ | $C_{11}$ | $C_{33}$ | $C_{44}$ | $C_{66}$ | $C_{12}$ | $C_{13}$ | $C^{P}_{1}$ | $C^{P}_{2}$ | Ref. |
|---|---|---|---|---|---|---|---|---|---|---|
| $Y_2GaB$ | 0 | 60.23 | 91.84 | 26.39 | 4.24 | 51.73 | 34.37 | 7.98 | 47.48 | This work |
| | | 90.83 | 118.95 | 36.14 | 18.48 | 53.87 | 41.97 | - | - | **[16]** |
| | 5 | 83.76 | 126.35 | 33.71 | 8.95 | 65.84 | 53.52 | 19.80 | 56.89 | This work |
| | 10 | 102.09 | 158.25 | 42.11 | 12.41 | 77.27 | 71.26 | 29.14 | 64.85 | This work |
| $Y_2InB$ | 0 | 78.05 | 124.01 | 33.08 | 9.87 | 58.30 | 36.30 | 3.21 | 48.43 | This work |
| | | 78.77 | 128.87 | 33.70 | 11.56 | 55.65 | 37.16 | - | - | **[16]** |

For elastically stable $Y_2GaB$, all displayed stiffness coefficients increase between 0 and 10 GPa. $C_{11}$ rises from 60.23 to 102.09 GPa and $C_{33}$ from 91.84 to 158.25 GPa, while $C_{44}$ and $C_{66}$ increase from 26.39 and 4.24 GPa to 42.11 and 12.41 GPa, respectively. The persistent relation $C_{33} > C_{11}$ shows that direct longitudinal deformation is more strongly resisted along *c*, whereas the much smaller $C_{66}$ identifies basal-plane shear as the softest deformation. At ambient pressure, $Y_2InB$ is stiffer than $Y_2GaB$. The larger shear constants reduce the pronounced basal-shear softness found in $Y_2GaB$. The $Y_2InB$ constants agree closely with the earlier calculation, whereas the present $Y_2GaB$ values are lower, particularly for $C_{11}$ and $C_{66}$ **[16]**. Related layered borides likewise show a marked separation between longitudinal and shear stiffness **[45,46]**.

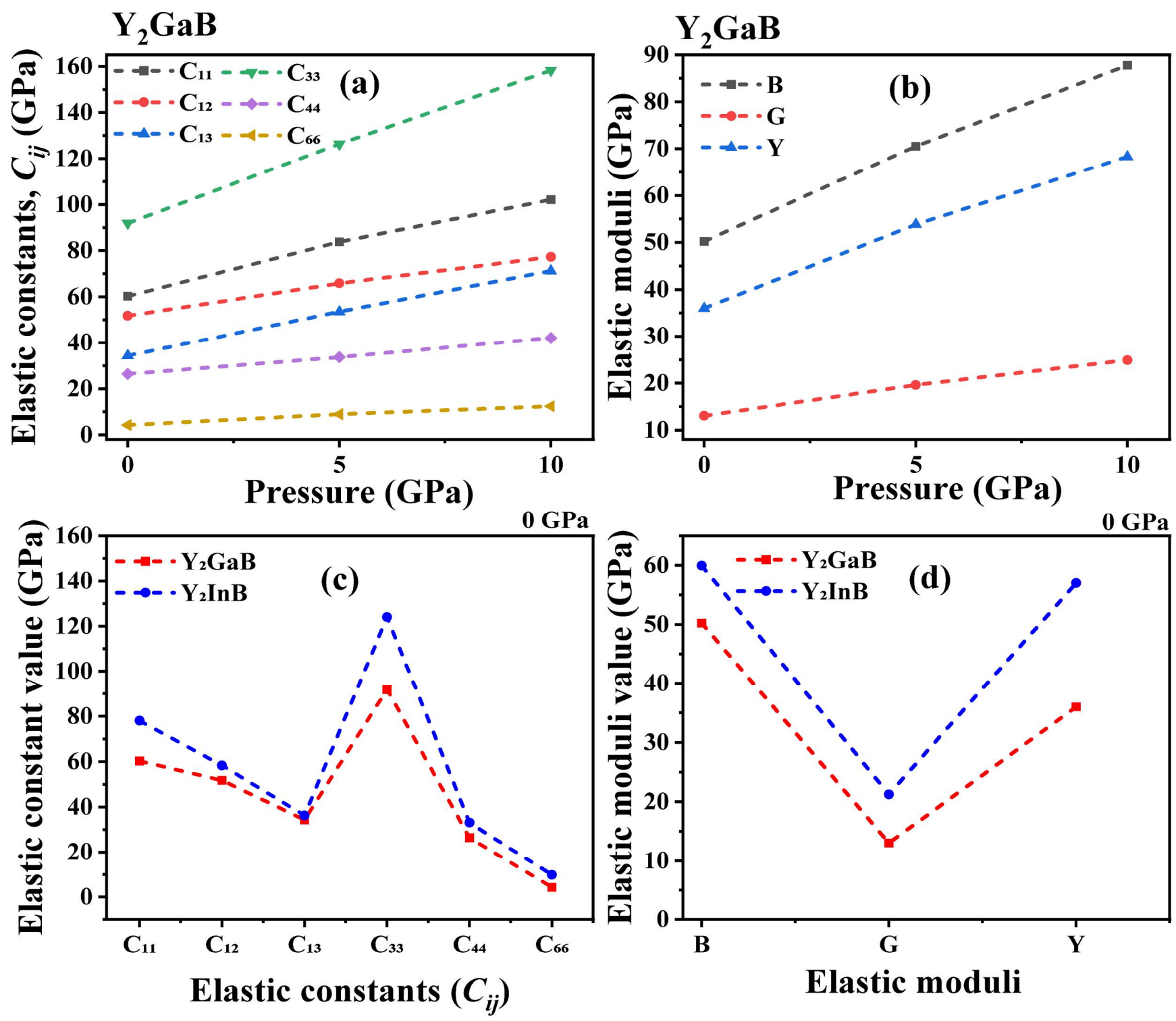

**Figure 3** (a) Pressure-dependent elastic constants and (b) Hill elastic moduli of $Y_2GaB$; ambient-pressure comparison of (*c*) the elastic constants and (d) the Hill elastic moduli of $Y_2GaB$ and $Y_2InB$.

The Hill bulk and shear moduli were taken as the arithmetic means of the Voigt and Reuss bounds **[47]**:

$$B_H = \frac{B_V + B_R}{2} \tag{7}$$

$$G_H = \frac{G_V + G_R}{2} \tag{8}$$

Young's modulus and Poisson's ratio were then obtained from:

$$Y = \frac{9B_H G_H}{3B_H + G_H} \tag{9}$$

$$\nu = \frac{3B_H - 2G_H}{2(3B_H + G_H)} \tag{10}$$

**Table 3** lists the polycrystalline moduli for the elastically stable states. In $Y_2GaB$, $B_H$ increases from 50.22 to 87.79 GPa between 0 and 10 GPa, while $G_H$ rises from 13.04 to 24.93 GPa and $Y$ from 36.01 to 68.32 GPa. Compression therefore increases the resistance to volumetric, shear, and uniaxial deformation throughout the stable pressure range. At 0 GPa, $Y_2InB$ has $B_H$ = 59.95 GPa, $G_H$ = 21.25 GPa, and $Y$ = 57.01 GPa, all above the corresponding $Y_2GaB$ values. This establishes the ambient-pressure compositional contrast without extending the comparison to elastically unstable $Y_2InB$ states. Both phases remain softer than several reported 212 borides, for which the additional B-containing layer is associated with larger calculated moduli [48,49]. Figure 3 illustrates different elastic parameters and their pressure evolution.

**Table 3** Calculated Voigt, Reuss, and Hill bulk moduli ($B_v$, $B_r$, and $B_h$), shear moduli ($G_V$, $G_R$, and $G_H$), and Hill Young's modulus ($Y$) for the elastically stable states. All values are in GPa; entries attributed to Ref. [16] are reproduced as reported.

| Compound | P (GPa) | $B_V$ | $B_R$ | $B_H$ | $G_V$ | $G_R$ | $G_H$ | $Y$ | Ref. |
|---|---|---|---|---|---|---|---|---|---|
| $Y_2GaB$ | 0 | 50.36 | 50.08 | 50.22 | 17.52 | 8.55 | 13.04 | 36.01 | This work |
| | | 64.03 | 63.75 | - | 29.01 | 26.09 | - | 72.27 | [16] |
| | 5 | 71.07 | 69.99 | 70.53 | 23.34 | 15.88 | 19.61 | 53.84 | |
| | 10 | 89.11 | 86.47 | 87.79 | 28.83 | 21.01 | 24.93 | 68.32 | This work |
| $Y_2InB$ | 0 | 60.21 | 59.69 | 59.95 | 25.15 | 17.34 | 21.25 | 57.01 | |
| | | 60.71 | 59.79 | - | 26.22 | 19.43 | - | 45.03 | [16] |

The derived indicators in Table 4 retain the same elastic-stability restriction. For $Y_2GaB$, $G_H/B_H$ changes only from 0.26 to 0.28 between 0 and 10 GPa, while $\nu$ remains within 0.37-0.38. Both lie on the ductile side of the conventional empirical limits $G/B$ = 0.57 and $\nu$ = 0.26 [50]. Ambient-pressure $Y_2InB$ shows the same modulus-based tendency, with $G_H/B_H$ = 0.35 and $\nu$ = 0.34. Together, these indicators consistently suggest a ductile elastic response for the studied compounds within their mechanically stable pressure range. These criteria do not predict plastic flow, fracture toughness, or failure under service loading. The universal anisotropy index $A^U$ is retained in Table 4 for continuity and is analyzed in Section 3.3 [20].

The two independent hexagonal Cauchy pressures are [51]

$$C_1^P = C_{13} - C_{44} \quad (11)$$
$$C_2^P = C_{12} - C_{66} \quad (12)$$

Both Cauchy pressures are positive for every elastically stable state. In $Y_2GaB$, $C_{13} - C_{44}$ increases from 7.98 to 29.14 GPa and $C_{12}$ - $C_{66}$ from 47.48 to 64.85 GPa between 0 and 10 GPa, whereas $Y_2InB$ gives 3.21 and 48.43 GPa at ambient pressure, respectively. Their signs are consistent with the ductile tendency inferred from the isotropic elastic moduli, although Cauchy pressure alone does not entirely establish bonding character or macroscopic ductility.

**Table 4** Calculated Pugh ratio $G_H/B_H$, Poisson's ratio $\nu$, machinability index $\mu^M = B_H/C_{44}$, Tian Vickers hardness $H_v$, and universal anisotropy index $A^U$ for the elastically stable states.

| Compound | P (GPa) | $G_H/B_H$ | $\nu$ | $\mu^M$ | $H_v$ (GPa) | $A^U$ | Ref. |
|---|---|---|---|---|---|---|---|
| $Y_2GaB$ | 0 | 0.26 | 0.38 | 1.90 | 1.22 | 5.25 | This work |
| | | 0.43 | 0.31 | -- | 1.349 | 0.67 | [16] |
| | 5 | 0.27 | 0.37 | 2.09 | 1.76 | 2.36 | |
| | 10 | 0.28 | 0.37 | 2.08 | 2.14 | 1.89 | This work |
| $Y_2InB$ | 0 | 0.35 | 0.34 | 1.81 | 2.46 | 2.26 | |
| | | 0.37 | 0.33 | -- | 1.24 | 1.77 | [16] |

The machinability index, $\mu^M$, and modulus-based Vickers hardness, $H_V$, were evaluated using **Eqs. (13) and (14)**, respectively **[52,53]**.

$$\mu^M = \frac{B_H}{C_{44}} \tag{13}$$

$$H_V = 0.92(\frac{G_H}{B_H})^{1.137} G_H^{0.708} \tag{14}$$

The machinability index of $Y_2GaB$ increases from 1.90 at 0 GPa to 2.09 at 5 GPa and remains nearly unchanged at 2.08 at 10 GPa, whereas $Y_2InB$ gives 1.81 at ambient pressure. These values indicate that both the MAB phases are fairly machinable and possess reasonable dry-lubricity.

Graphical representations of these derived mechanical parameters are given in **Figures 4-5**.

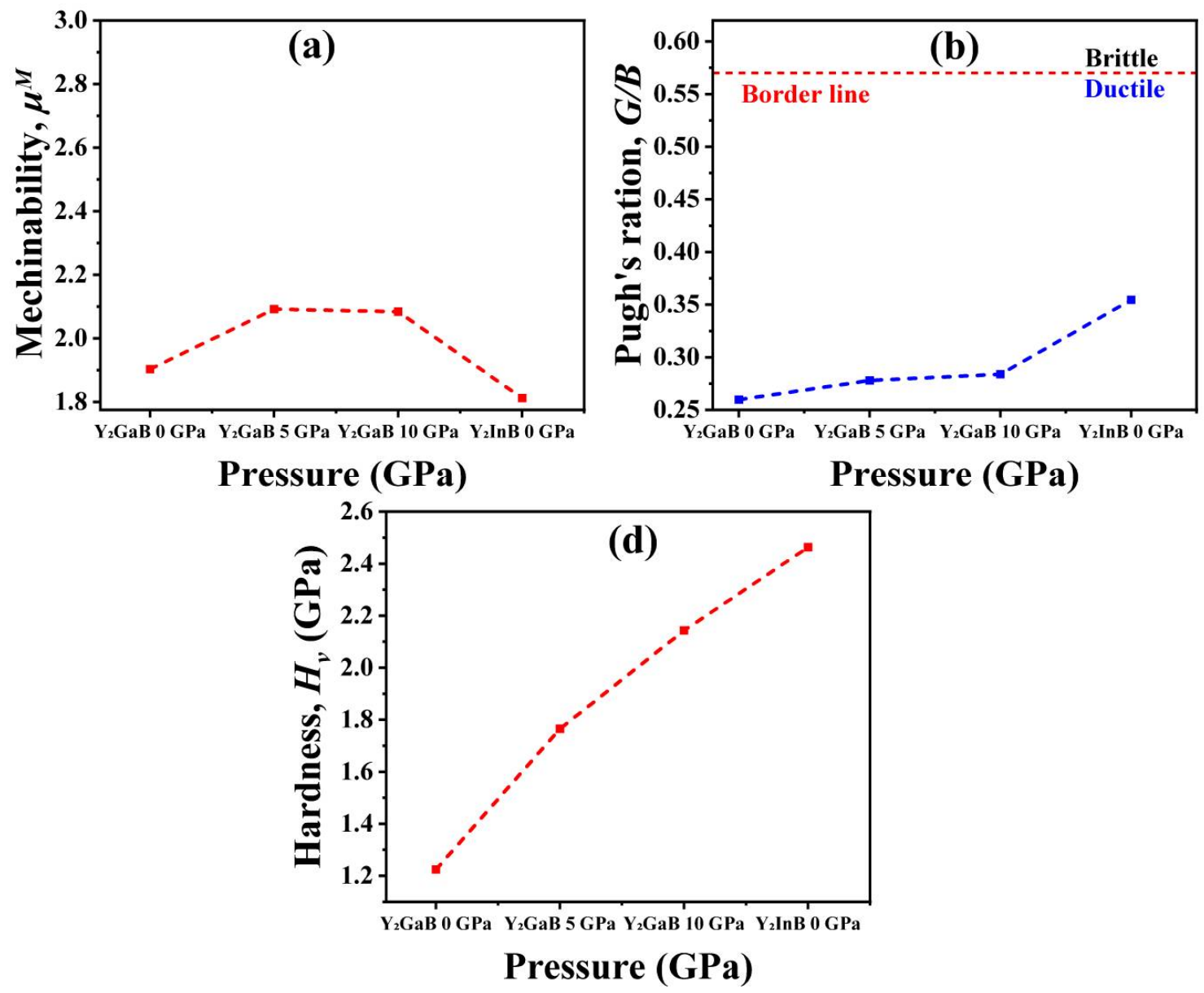

**Figure 4** Mechanical indicators for the elastically stable states: (a) machinability index, (b) Pugh ratio, and (d) Vickers hardness. Dashed lines at *G/B* = 0.57 mark the conventional brittle–ductile boundaries.

The Tian-model hardness of $Y_2GaB$ rises from 1.22 to 2.14 GPa across its elastic-stability range, while $Y_2InB$ gives 2.46 GPa at 0 GPa. The low values characterize both compounds as mechanically soft within this empirical model and should not be interpreted as indentation measurements **[53,46,49]**. The elastically stable dataset provides the basis for the directional analysis in Section 3.3.

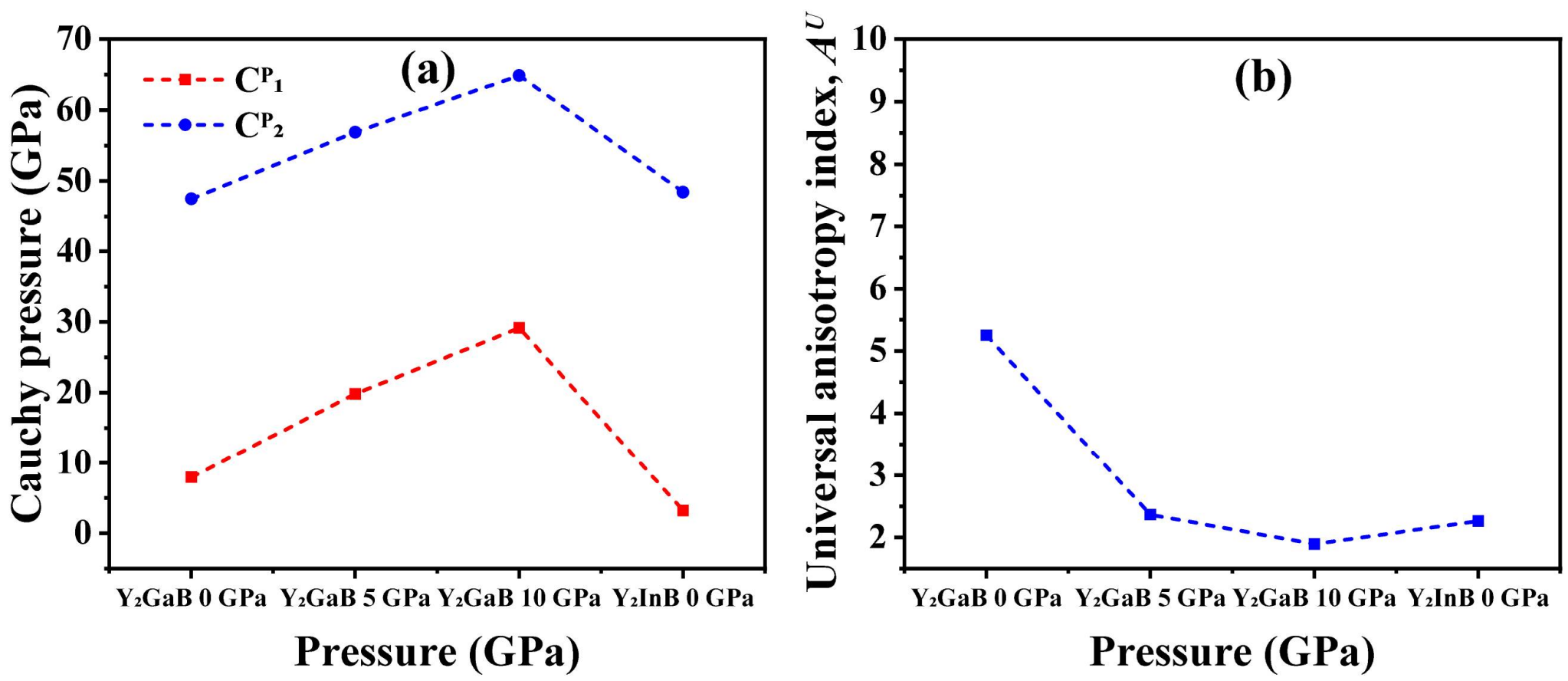

**Figure 5** Calculated (a) Cauchy pressures $C_{13} - C_{44}$ and $C_{12} - C_{66}$ and (b) universal elastic anisotropy index $A^U$ for the elastically stable states of $Y_2GaB$ and $Y_2InB$.

### 3.3 Elastic Anisotropy

Overall elastic anisotropy was quantified using the universal index $A^U$, which compares the Voigt and Reuss bounds [20]:

$$A^U = 5\frac{G_V}{G_R} + \frac{B_V}{B_R} - 6 \tag{15}$$

where $B_V$ and $B_R$ are the Voigt and Reuss bulk moduli, and $G_V$ and $G_R$ are the corresponding shear moduli. $A^U = 0$ signifies isotropy, whereas larger values indicate greater aggregate contrast [54,55]. For $Y_2GaB$, $A^U$ decreases from 5.25 at 0 GPa to 2.36 at 5 GPa and 1.89 at 10 GPa (**Table 4**), showing that compression reduces the overall elastic anisotropy across the elastically stable range. Ambient-pressure $Y_2InB$ gives $A^U = 2.26$, below $Y_2GaB$ at 0 GPa but slightly above compressed $Y_2GaB$. The high initial value for $Y_2GaB$ is dominated by the large Voigt-Reuss separation in shear, consistent with its small basal shear stiffness $C_{66}$.

Directional Young's modulus, shear modulus, linear compressibility, and Poisson's ratio were evaluated from the elastic tensor and visualized as two-dimensional sections and three-dimensional surfaces [56,57]. These representations resolve property-specific directions that cannot be inferred from $A^U$ alone: a sphere is the isotropic reference, whereas departures from sphericity reflect directional variation. The results of this analysis are shown in **Figures 6-13**.

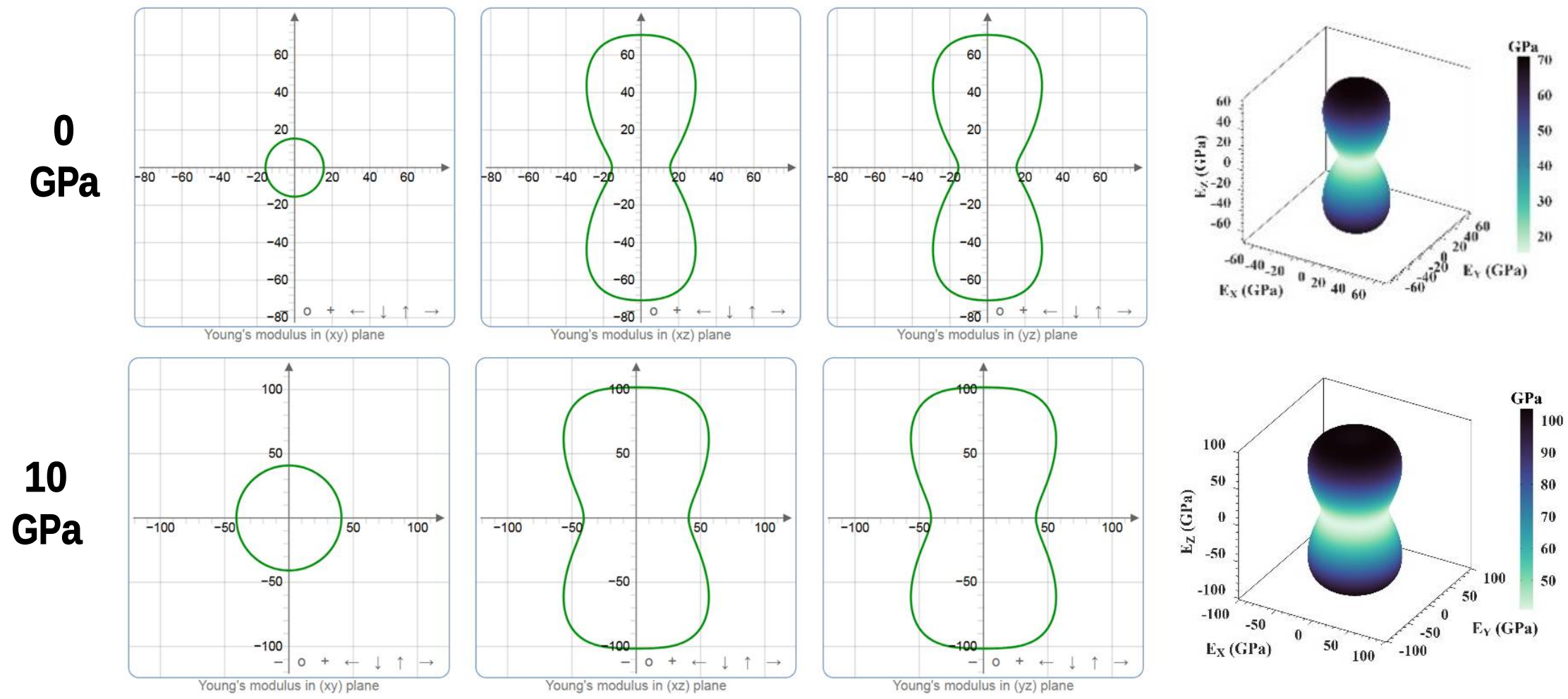

**Figure 6** Directional variation in Young's modulus $Y$ of $Y_2GaB$ at 0 and 10 GPa, shown as two-dimensional sections in the *xy*-, *xz*-, and *yz*-planes and as three-dimensional surfaces.

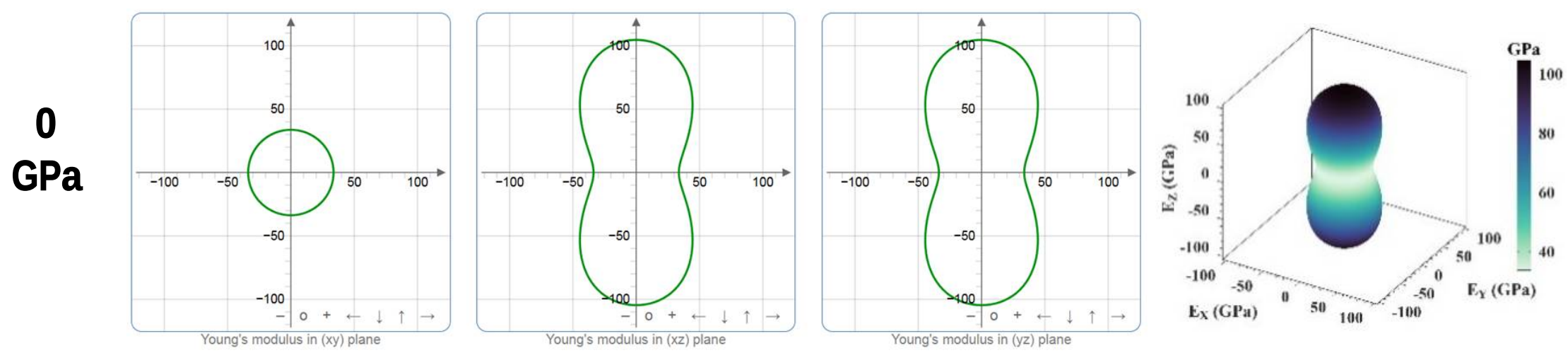

**Figure 7** Directional variation in Young's modulus $Y$ of $Y_2InB$ at 0 GPa, shown as two-dimensional sections in the *xy*-, *xz*-, and *yz*-planes and as a three-dimensional surface.

At 0 GPa, the Young's modulus of $Y_2GaB$ spans 15.47-70.74 GPa ($Y_{max}/Y_{min}$ = 4.57), with the largest stiffness along *c* and a much softer basal response. At 10 GPa, the range shifts to 40.85–103.64 GPa and the ratio falls to 2.54. All directions stiffen, but the initially compliant directions strengthen proportionally more, consistent with the decline in $A^U$.

Ambient-pressure $Y_2InB$ spans 33.71-104.68 GPa ($Y_{max}/Y_{min}$ = 3.10). Its minimum and maximum exceed those of $Y_2GaB$ at the same pressure, while the smaller ratio reflects a stronger basal response. The maximum remains aligned with *c*, preserving the basal-axial contrast expected for the hexagonal structure.

The principal linear compressibilities of a hexagonal crystal are:

$$\beta_a = \frac{C_{33} - C_{13}}{D} \quad (16)$$

$$\beta_c = \frac{C_{11} + C_{12} - 2C_{13}}{D} \quad (17)$$

with

$$D = (C_{11} + C_{12})\,C_{33} - 2{C_{13}}^2 \quad (18)$$

where $\beta_a$ and $\beta_c$ are the basal-plane and *c*-axis compressibilities, respectively **[58,59]**.

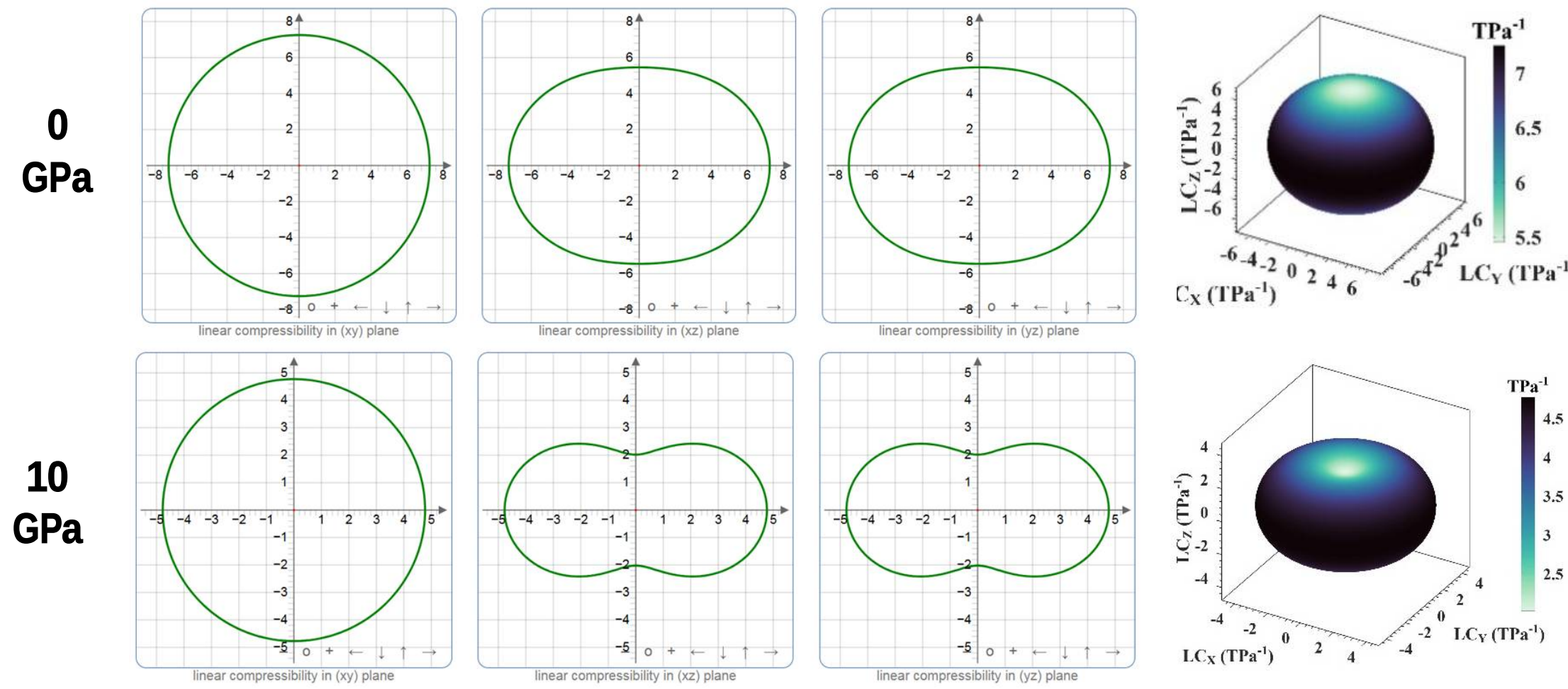

**Figure 8** Directional variation in linear compressibility $\beta$ of $Y_2GaB$ at 0 and 10 GPa, shown as two-dimensional sections in the *xy*-, *xz*-, and *yz*-planes and as three-dimensional surfaces.

For $Y_2GaB$, $\beta_a$ and $\beta_c$ decrease from $7.257 \times 10^{-3}$ and $5.457 \times 10^{-3}$ $GPa^{-1}$ at 0 GPa to $4.772 \times 10^{-3}$ and $2.021 \times 10^{-3}$ $GPa^{-1}$ at 10 GPa. Both directions become less compressible, but $\beta_a/\beta_c$ rises from 1.33 to 2.36 because $\beta_c$ decreases more rapidly. The lower $A^U$ therefore does not imply that every directional contrast diminishes: tensile and shear anisotropy decrease, whereas the basal-to-axial compressibility contrast increases.

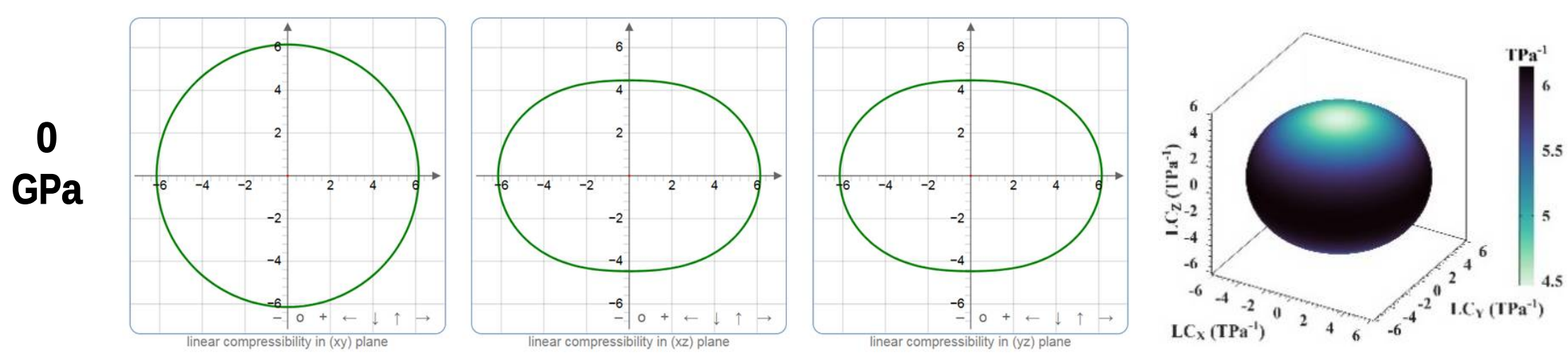

**Figure 9** Directional variation in linear compressibility $\beta$ of $Y_2InB$ at 0 GPa, shown as two-dimensional sections in the *xy*-, *xz*-, and *yz*-planes and as a three-dimensional surface.

At 0 GPa, $Y_2InB$ has $\beta_a = 6.144 \times 10^{-3}$ and $\beta_c = 4.466 \times 10^{-3}$ $GPa^{-1}$, giving $\beta_a/\beta_c = 1.38$. Both values are below those of $Y_2GaB$, but the common relation $\beta_a > \beta_c$ shows that Ga-to-In substitution retains the same directional compression pattern.

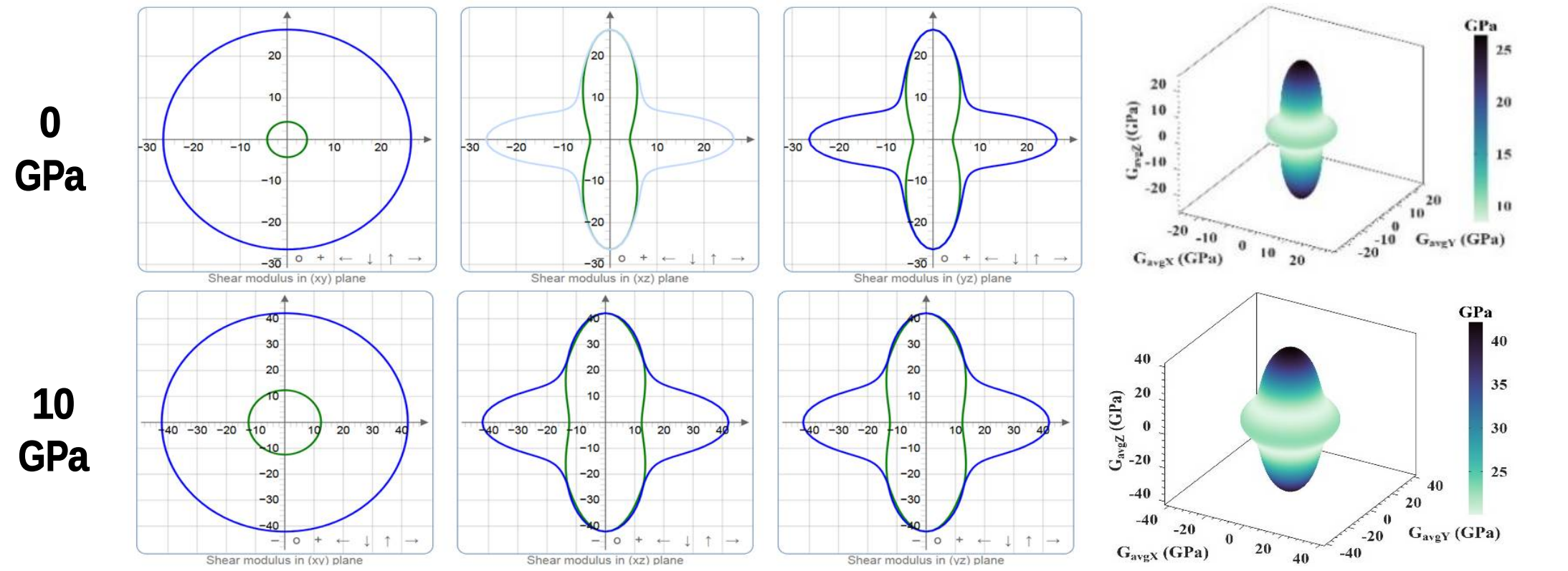

**Figure 10** Directional variation in shear modulus *G* of $Y_2GaB$ at 0 and 10 GPa, shown as two-dimensional sections in the *xy*-, *xz*-, and *yz*-planes and as three-dimensional surfaces.

Shear is the most anisotropic response of ambient-pressure $Y_2GaB$: *G* ranges from 4.24 to 26.39 GPa ($G_{max}/G_{min}$ = 6.22). At 10 GPa, the extrema increase to 12.41 and 42.11 GPa and the ratio falls to 3.39. The weakest shear mode therefore stiffens faster than the strongest, accounting for much of the reduction in the aggregate index.

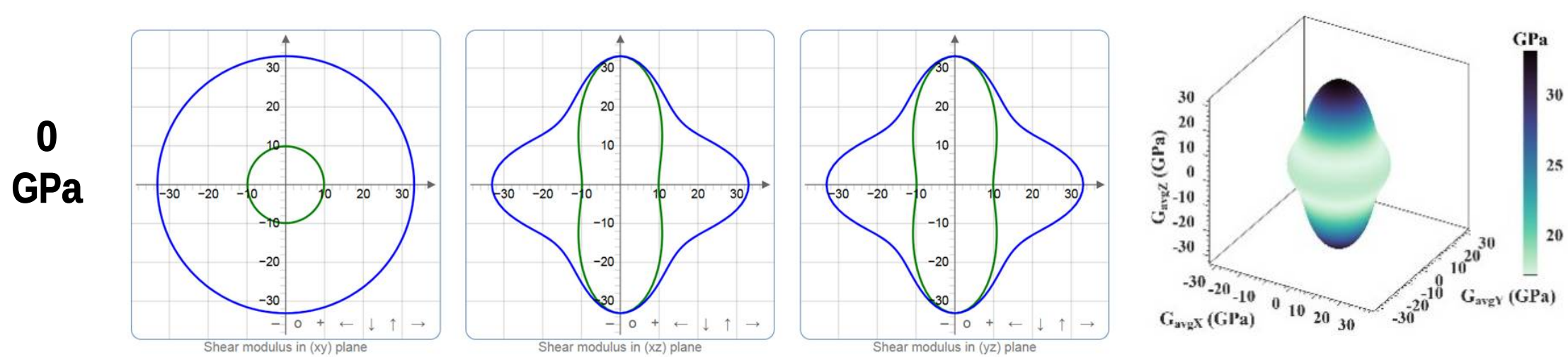

**Figure 11** Directional variation in shear modulus *G* of $Y_2InB$ at 0 GPa, shown as two-dimensional sections in the *xy*-, *xz*-, and *yz*-planes and as a three-dimensional surface.

$Y_2InB$ gives $G_{min}$ = 9.87 and $G_{max}$ =33.08 GPa ($G_{max}/G_{min}$ = 3.35) at 0 GPa. Its minimum shear resistance is more than twice that of ambient-pressure $Y_2GaB$, whereas the maximum is only moderately higher. The resulting directional contrast is close to that of $Y_2GaB$ at 10 GPa.

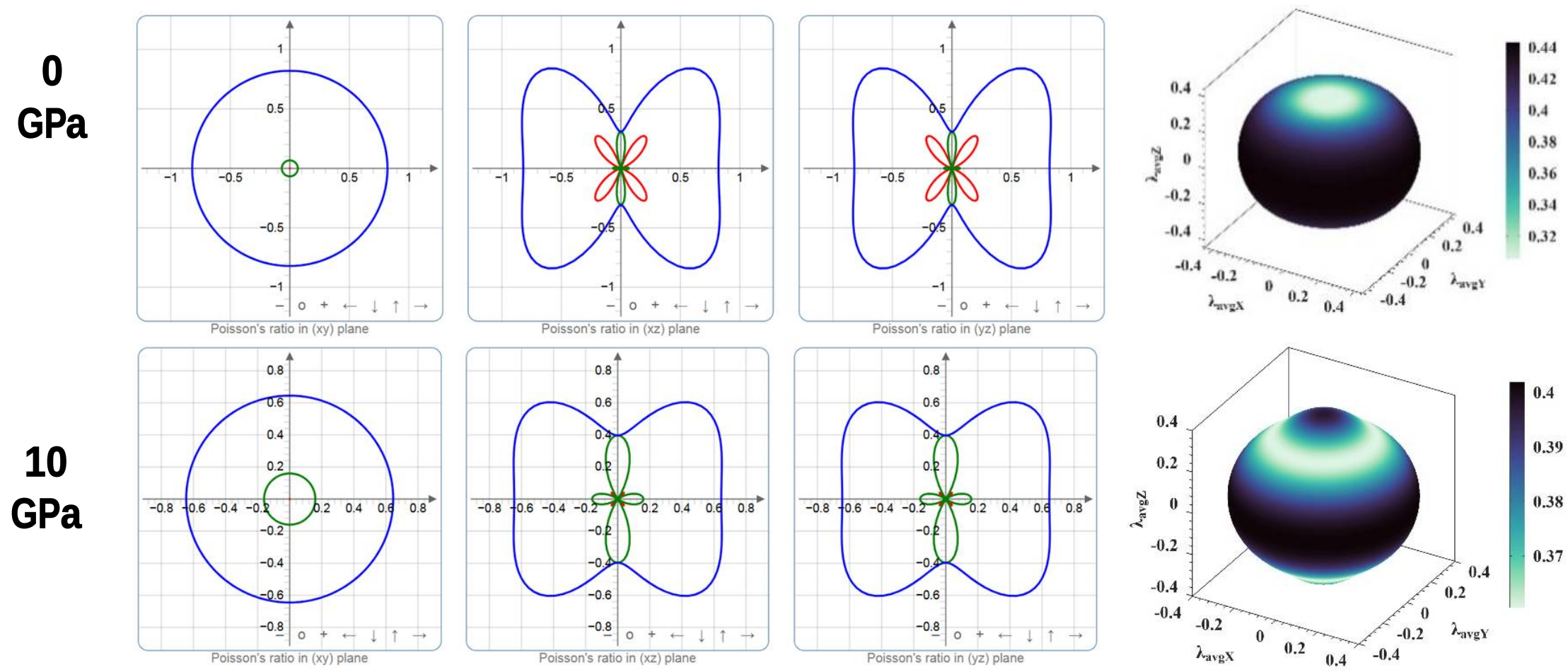

**Figure 12** Directional variation in Poisson's ratio $v$ of $Y_2GaB$ at 0 and 10 GPa, showing the minimum and maximum transverse responses in the principal crystallographic sections and the corresponding three-dimensional surfaces.

Directional Poisson's ratio depends on both the loading and transverse directions, so anisotropic values may be negative or exceed 0.5 without violating elastic positive definiteness **[58–61]**. For $Y_2GaB$, the range narrows from -0.339($v_{min}$) -1.068($v_{max}$) at 0 GPa to -0.056($v_{min}$)-0.7859( $v_{max}$) at 10 GPa. Negative values occur only for selected orientation pairs and indicate a restricted directional auxetic response rather than isotropic auxeticity.

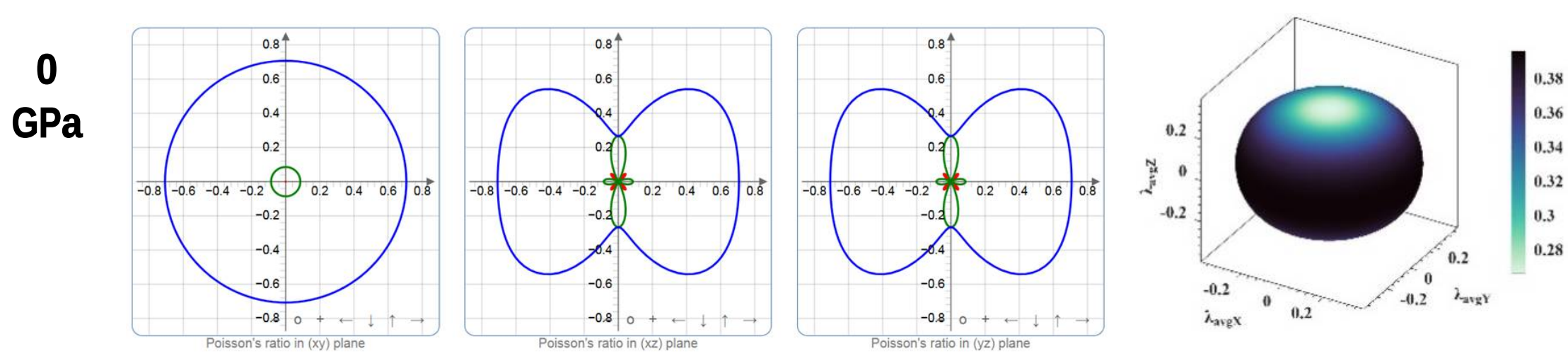

**Figure 13** Directional variation in Poisson's ratio $v$ of $Y_2InB$ at 0 GPa, showing the minimum and maximum transverse responses in the principal crystallographic sections and the corresponding three-dimensional surfaces.

Ambient-pressure $Y_2InB$ spans -0.062($v_{min}$) -0.748($v_{max}$), a markedly narrower interval than $Y_2GaB$ at 0 GPa and similar to compressed $Y_2GaB$. Across the directional surfaces, pressure and In substitution reduce the exceptional tensile and shear contrasts of ambient-pressure $Y_2GaB$, while noticeable non-sphericity and basal-axial differentiation remain **[62,63]**. The elastically stable dataset provides the basis for the thermophysical descriptors in Section 3.4.

## 3.4 Thermophysical Properties

The thermophysical analysis was confined to the elastically stable states identified in Section 3.2, $Y_2GaB$ at 0, 5, and 10 GPa and $Y_2InB$ at 0 GPa. Except for the Slack estimate at 300 K, all quantities were derived from static-lattice structural and elastic data and therefore correspond approximately to 0 K. The acoustic velocities, elastic Debye temperature, empirical melting-temperature estimate, and Clarke minimum phonon thermal conductivity describe the isotropic Hill aggregate rather than direction-resolved propagation in a single crystal.

### 3.4.1 Acoustic Velocities and Debye Temperature

The longitudinal, transverse, and mean acoustic velocities were obtained from the Hill bulk modulus $B_H$, Hill shear modulus $G_H$, and density $\rho$ **[47,64]**:

$$v_l = \left[\frac{B_H + \frac{4G_H}{3}}{\rho}\right]^{\frac{1}{2}} \tag{19}$$

$$v_t = \left(\frac{G_H}{\rho}\right)^{\frac{1}{2}} \tag{20}$$

$$v_m = \left[\frac{1}{3}\left(\frac{2}{v_t^{\,3}} + \frac{1}{v_l^{\,3}}\right)\right]^{\frac{-1}{3}} \tag{21}$$

Here $v_l$, $v_t$, and $v_m$ denote the longitudinal, transverse, and mean sound velocities, respectively. $B_H$ and $G_H$ are expressed in pascals, and $\rho$ in kg m$^{-3}$. The elastic-acoustic Debye temperature was then calculated using the Anderson relation **[64]**:

$$\Theta_D = \frac{h}{k_B}\left[\frac{3n_f N_A \rho}{4\pi M}\right]^{\frac{1}{3}} v_m \tag{22}$$

In **Eq. (22)**, $h$ is Planck's constant, $k_B$ is the Boltzmann constant, $n_f = 4$ is the number of atoms in one $Y_2AB$ formula unit, $N_A$ is Avogadro's constant, and $M$ is the molar mass of one formula unit in kg mol$^{-1}$; $\rho$ is entered in kg m$^{-3}$. The resulting $\Theta_D$ is therefore an elastic-acoustic estimate rather than a value derived from the phonon DOS.

Under compression, $Y_2GaB$ exhibits a clear enhancement in its acoustic properties, as reported in **Table 5**. The longitudinal ($v_l$), transverse ($v_t$), and average ($v_m$) sound velocities increase progressively with pressure, indicating a gradual stiffening of the lattice. A similar pressure-induced increase is observed in the Debye temperature ($\Theta_D$), reflecting stronger interatomic interactions and an overall increase in lattice rigidity. At ambient pressure, $Y_2InB$ also shows relatively strong acoustic characteristics, as evidenced by its sound velocities and Debye temperature. Overall, these results demonstrate that hydrostatic pressure strengthens the elastic and vibrational response of $Y_2GaB$. This contrast follows from its larger elastic moduli despite the higher density. Similar pressure-driven stiffening has been reported for related layered borides **[45,48]**.

### 3.4.2 Estimated Melting Temperature

The melting temperature was estimated from an empirical correlation with the longitudinal elastic constants **[65]**:

$$T_m = 354 + 4.5\left(\frac{2C_{11} + C_{33}}{3}\right) \tag{23}$$

With $C_{11}$ and $C_{33}$ expressed in GPa, **Eq. (23)** gives a central estimate. For $Y_2GaB$, the estimate rises from 672.45 K at 0 GPa to 794.81 K at 5 GPa and 897.64 K at 10 GPa, following the increase in the longitudinal stiffnesses. Ambient-pressure $Y_2InB$ gives 774.16 K, higher than $Y_2GaB$ at 0 GPa, but lower than the values obtained for compressed $Y_2GaB$. Relatively low values of melting temperatures of the two compounds follow from their soft nature as a result of weak interatomic bondings.

### 3.4.3 Minimum Thermal Conductivity

The phenomenological lower bound to thermal conductivity was estimated with the Clarke model [66]:

$$k_{min} = k_B v_m \left[\frac{M}{n_f \rho N_A}\right]^{\frac{-2}{3}} \quad (24)$$

In Eq. (24), $k_{min}$ is the Clarke minimum thermal conductivity, while $M/(n_f \rho N_A)$ represents the average atomic volume, with $n_f = 4$. The resulting thermophysical quantities are collected in **Table 5**.

**Table 5** Elastic-derived thermophysical descriptors of the elastically stable states of $Y_2GaB$ and $Y_2InB$. The listed melting temperatures are the central values from Eq. (23).

| Compound | $P$ (GPa) | $\rho$ (g cm$^{-3}$) | $v_l$ (m s$^{-1}$) | $v_t$ (m s$^{-1}$) | $v_m$ (m s$^{-1}$) | $\Theta_D$ (K) | Estimated $T_m$ (K) | Clarke $k_{min}$ (W m$^{-1}$ K$^{-1}$) | Ref. |
|---|---|---|---|---|---|---|---|---|---|
| $Y_2GaB$ | 0 | 4.59 | 3837.86 | 1685.58 | 1903.01 | 198.18 | 672.45 | 0.321 | This work |
| | 5 | 4.98 | 4406.11 | 1984.38 | 2237.98 | 239.49 | 794.81 | 0.399 | |
| | 10 | 5.32 | 4769.64 | 2164.61 | 2440.42 | 266.96 | 897.64 | 0.455 | |
| $Y_2InB$ | 0 | 5.16 | 4136.37 | 2029.34 | 2279.01 | 233.89 | 774.16 | 0.374 | |

The Clarke estimate follows the same pressure trend as the acoustic velocities. For $Y_2GaB$, $k_{min}$ increases from 0.321 W m$^{-1}$ K$^{-1}$ at 0 GPa to 0.399 W m$^{-1}$ K$^{-1}$ at 5 GPa and 0.455 W m$^{-1}$ K$^{-1}$ at 10 GPa, as the mean acoustic velocity rises and the atomic volume contracts. Ambient-pressure $Y_2InB$ gives 0.374 W m$^{-1}$ K$^{-1}$, above $Y_2GaB$ at 0 GPa but below the two compressed states. These lower-bound estimates are very low, suggesting that high temperature thermal conductivity of the MAB compounds under consideration are quite small.

### 3.4.4 Slack-Estimated Lattice Thermal Conductivity

To complement the lower-bound estimate, the lattice thermal conductivity was evaluated at $T = 300$ K using the Slack model [67]:

$$\kappa_L(T) = \frac{A(\gamma)\overline{M}{\Theta_D}^3\delta}{\gamma^2 n_c^{\frac{2}{3}} T} \quad (25)$$

Here $\overline{M} = M_{formula}/4$ is the average atomic molar mass, $\delta = (V_{cell}/n_c)^{1/3}$ is the cube root of the atomic volume, and $n_c = 8$ is the number of atoms in the $P6_3/mmc$ crystallographic cell. $\Theta_D$ is the elastic Debye temperature. Using $\overline{M}$ in kg mol$^{-1}$, $\delta$ in m, and $\Theta_D$ and $T$ in K gives $\kappa_L$ in W m$^{-1}$ K$^{-1}$. The acoustic Grüneisen parameter was inferred from the Poisson ratio $\nu$ [68], and the $\gamma$-dependent Julian coefficient was evaluated as [69]:

$$\gamma = \frac{3(1+\nu)}{2(2-3\nu)} \quad (26)$$

$$A(\gamma) = \frac{5.720 \times 10^7 \times 0.849}{2\left[1 - \frac{0.514}{\gamma} + \frac{0.228}{\gamma^2}\right]} \quad (27)$$

**Table 6** summarizes the calculated lattice thermal conductivities.

**Table 6** Poisson ratio, estimated Grüneisen parameter, and Slack-estimated lattice thermal conductivity at 300 K for the elastically stable states.

| Compound | ***P*** (GPa) | ***ν*** | Estimated ***γ*** | Slack $\kappa_L$ at 300 K ($W\ m^{-1}\ K^{-1}$) | Ref. |
|---|---|---|---|---|---|
| $Y_2GaB$ | 0 | 0.3805 | 2.4121 | 0.605 | This work |
| | 5 | 0.3728 | 2.3356 | 1.114 | |
| | 10 | 0.3703 | 2.3119 | 1.543 | |
| $Y_2InB$ | 0 | 0.3415 | 2.0628 | 1.664 | |

At 300 K, the Slack estimate for $Y_2GaB$ increases from 0.605 W m$^{-1}$ K$^{-1}$ at 0 GPa to 1.114 W m$^{-1}$ K$^{-1}$ at 5 GPa and 1.543 W m$^{-1}$ K$^{-1}$ at 10 GPa. The dominant contribution comes from the $\Theta_D^3$ term and the modest fall in $\gamma$ further raises $\kappa_L$, whereas the reduction in $\delta$ partly offsets the increase. $Y_2InB$ gives 1.664 W m$^{-1}$ K$^{-1}$ at ambient pressure, slightly above $Y_2GaB$ at 10 GPa. Its lower estimated Grüneisen parameter compensates for the lower Debye temperature when combined with the larger average atomic mass and atomic-volume factor. These differences are best viewed as comparative room-temperature model estimates, not as precise transport predictions.

The limits of this comparison are important. The Slack model is primarily a high-temperature approximation, and here $\gamma$ is inferred from an isotropic Hill Poisson ratio rather than obtained from mode-resolved anharmonic calculations. Third-order interatomic force constants and anharmonic Boltzmann transport were not evaluated, so defect, boundary, isotope, electron–phonon, and explicit mode-resolved scattering are absent. The Slack values should therefore be distinguished from both the Clarke lower bounds and experimentally measured total thermal conductivity, which would also contain an electronic contribution because $Y_2GaB$ and $Y_2InB$ are metallic.

## 3.5 Electronic Band Structure and Spin-Orbit Coupling

The electronic band structures of $Y_2GaB$ at 0 and 10 GPa and $Y_2InB$ at 0 GPa were calculated along the path Γ–A–H–K–Γ–M–L–H in the first Brillouin zone (**Figure 14**). Several bands cross $E_F$ in every case, and no continuous gap separates the occupied and unoccupied manifolds. Both compounds are therefore multiband metals in the states considered, consistent with earlier calculations for related layered $M_2AB$ and $M_2GaB$ borides **[13,40]**. At ambient pressure, nine $Y_2GaB$ bands intersect $E_F$. Crossings occur along Γ–A, A–H, K–Γ, Γ–M, M–L, and L–H, whereas H–K contains no Fermi-level intersection. The near-$E_F$ manifold contains both weakly and strongly dispersive bands. This variation reflects directional differences in band dispersion and charge carrier effective mass related transport anisotropy.

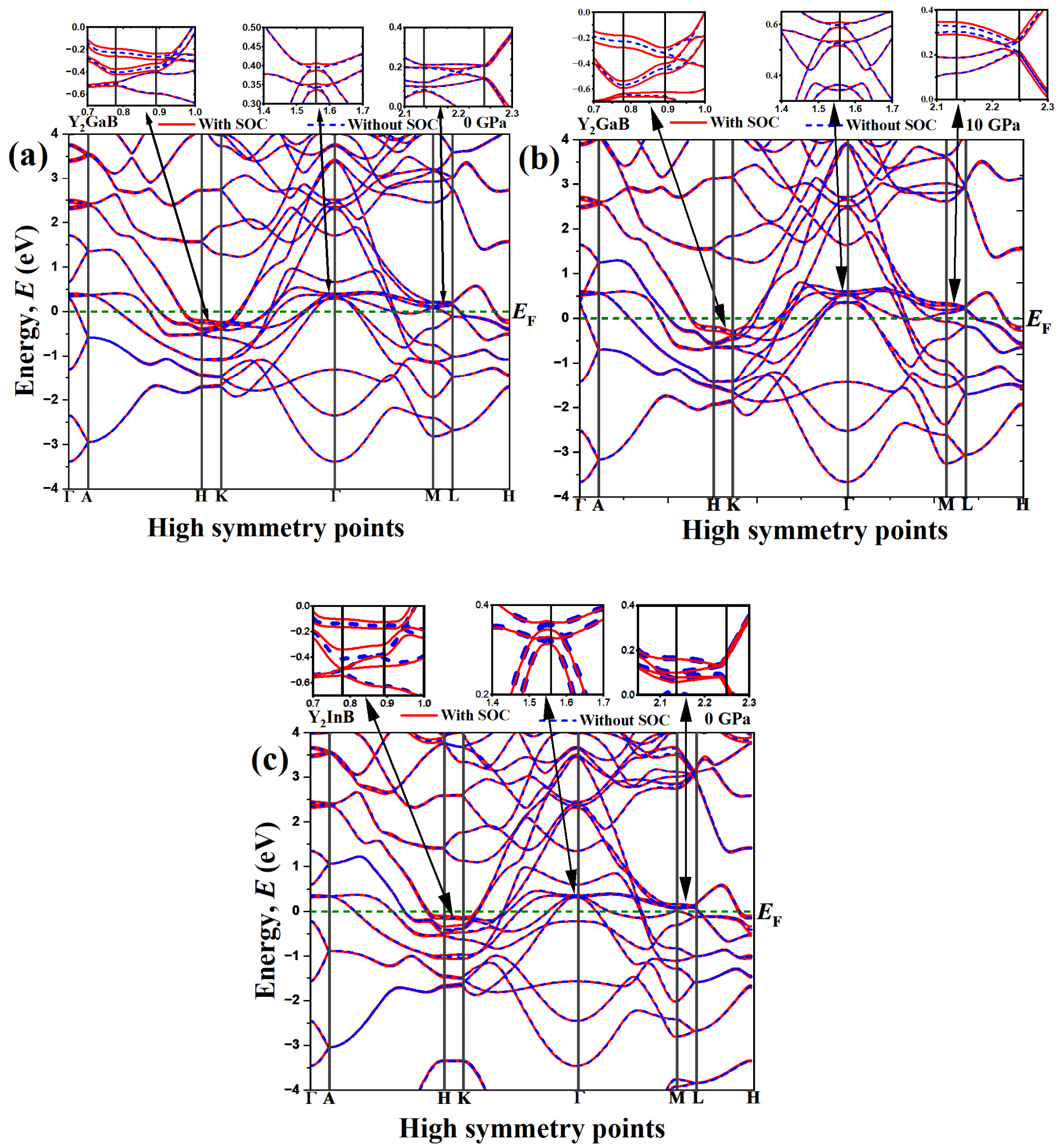

**Figure 14** Electronic band structures of (a) $Y_2GaB$ at 0 GPa, (b) $Y_2GaB$ at 10 GPa, and (c) $Y_2InB$ at 0 GPa, calculated with SOC (red solid lines) and without SOC (blue dashed lines) along Γ–A–H–K–Γ–M–L–H. The Fermi level is set to $E_F$ = 0 eV; the insets enlarge the near-$E_F$ regions.

At 10 GPa, $Y_2GaB$ still has nine bands crossing $E_F$ and the same path segments retain intersections. Compression therefore preserves the multiband metallic state while shifting individual bands weakly and nonuniformly. At M, the closest non-SOC state moves from approximately 0.086 eV at 0 GPa to -0.071 eV at 10 GPa, while states near A, H, K, and L undergo smaller shifts in different directions. The pressure response is consequently non-rigid: nearby extrema and crossings are rearranged rather than translated together.

Ambient-pressure $Y_2InB$ has eight bands crossing $E_F$, one fewer than $Y_2GaB$. Intersections occur along A–H, K–Γ, Γ–M, M–L, and L–H, with none along Γ–A or H–K. A non-SOC state at M lies only about 0.012 eV above $E_F$, but it does not add another crossing. The absent Γ–A intersection is the clearest near-Fermi distinction from $Y_2GaB$ and is consistent with the different Fermi-surface topology discussed in Section 3.7.

SOC modifies selected dispersions without opening any global gap or changing the number of bands that cross $E_F$. $Y_2GaB$ retains nine bands at both pressures, while $Y_2InB$ retains eight bands at 0 GPa. Because inversion and time-reversal symmetries are preserved, the SOC bands remain Kramers degenerate. The relativistic effect therefore appears mainly as weak local energy shifts and band rearrangements, not as a resolved spin splitting of each Kramers pair. Most changes near $E_F$ are modest, although selected $Y_2InB$ bands show more visible shifts. At M, for example, the $Y_2InB$ state moves from about 0.012 eV without SOC to 0.003 eV with SOC, without producing an additional crossing. These local changes do not establish band inversion, topological character, Rashba splitting, or a Lifshitz transition. The orbital character of the near-$E_F$ states and their reciprocal-space consequences are examined through the DOS and Fermi-surface analyses in Sections 3.6 and 3.7. Overall, the effect of SOC on the band structure is not significant.

## 3.6 Total and Partial Density of States

**Figure 15** compares the total and orbital-projected partial densities of states (PDOS) of $Y_2GaB$ at 0 and 10 GPa with those of $Y_2InB$ at 0 GPa. All three states retain a finite total density of states (TDOS) at $E_F$, confirming the multiband metallic character identified from the band structures. The near-$E_F$ spectrum is dominated by Y-4*d* states, with secondary contributions from B-2*p*, Ga-4*p*, and In-5*p*. The Y-5*s*, Y-4*p*, B-2*s*, and A-site *s* components are comparatively weak. Overlapping *d* and *p* features across the occupied manifold and immediately above $E_F$ show that several sublattices contribute within the same energy ranges. Similar spectra have been reported for $Y_2AB$ and related $M_2AB$ and $M_2GaB$ borides **[13,16,40,41]**. Such overlap supports orbital mixing and covalent bonding.

Compression reshapes the $Y_2GaB$ spectrum without removing the finite TDOS at $E_F$. The principal occupied feature moves from approximately −0.34 eV at 0 GPa to −0.25 eV at 10 GPa, whereas the feature near −1.07 eV shifts to about −1.22 eV. Their opposite movements rule out a simple rigid translation of the spectrum. The Y-4*d* contribution remains dominant near $E_F$, but the B-2*p* weight at the Fermi level decreases more visibly than the Ga-4*p* component, while the total DOS changes only modestly. Pressure therefore redistributes the relative positions and intensities of the orbital manifolds while preserving the metallic state. These spectral changes alone do not establish an electronic phase transition.

Ambient-pressure $Y_2InB$ retains the same orbital hierarchy, although the shape and placement of its main features differ from those of $Y_2GaB$. The largest contribution at $E_F$ again comes from Y-4*d* states, accompanied by In-5*p* and B-2*p*. Relative to $Y_2GaB$, the Y *d* contribution is stronger and the near-Fermi B *p* component is reduced. Its principal occupied feature lies near −0.25 eV, with a broader mixed feature around −2.03 eV. These differences reflect the combined influence of A-site orbital energies, local orbital mixing, and lattice geometry rather than atomic mass alone. The TDOS and PDOS shown here were calculated without SOC, so relativistic changes in the orbital-resolved spectra are not assessed in this section. The corresponding Fermi-surface geometry is examined next in Section 3.7, while the bonding interpretation is reserved for Section 3.8.

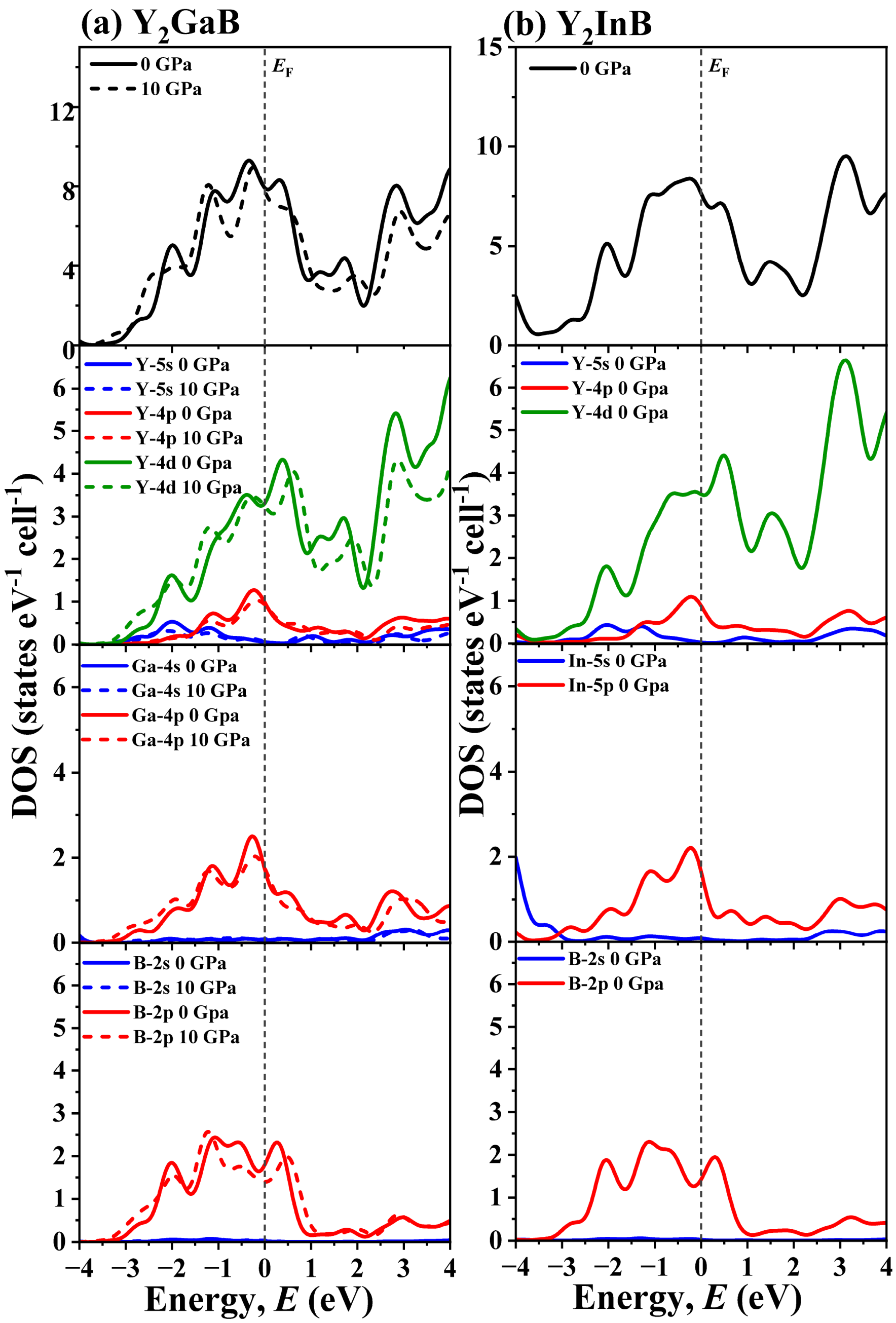

**Figure 15** TDOS and orbital-resolved PDOS of (a) $Y_2GaB$ at 0 GPa (solid) and 10 GPa (dashed) and (b) $Y_2InB$ at 0 GPa, calculated without SOC. $E_F$ = 0 eV; DOS is in states $eV^{-1}$ $cell^{-1}$.

## 3.7 Fermi-Surface Topology

The calculated Fermi surfaces provide the reciprocal-space counterpart to the multiband crossings and finite DOS discussed in Sections 3.5 and 3.6. They were constructed from the non-SOC bands using a dense 45 × 45 × 9 k-point mesh at $E_F$ = 0 eV for $Y_2GaB$ at 0 and 10 GPa and for $Y_2InB$ at 0 GPa. Nine band-resolved sheets are present in $Y_2GaB$ at both pressures, whereas $Y_2InB$ contains eight. At ambient pressure, $Y_2GaB$ displays a compact central pocket, several toroidal or strongly corrugated closed sheets, extended components that reach the Brillouin-zone boundary and small peripheral hole pockets. Layered MAX phases likewise exhibit mixtures of tubular, corrugated, and multiply connected Fermi-surface sheets **[70–73]**.

Compression to 10 GPa retains all nine $Y_2GaB$ sheets. The central pocket, surrounding rings, corrugated outer surfaces, and boundary pockets remain recognizable, although their dimensions, curvature, and warping change. Pressure therefore reshapes the existing sheets without reducing the multiband character. No sheet appears or disappears, and no clear change in connectivity is resolved between the two sampled pressures. The endpoint comparison consequently provides no evidence of a Lifshitz transition, which requires an actual change in Fermi-surface topology **[74]**. The small Fermi sheet at the center of each rectangular face shrink significantly at 10 GPa. Thus, a Lifshitz transition at an unsampled pressure cannot, however, be ruled out for $Y_2GaB$.

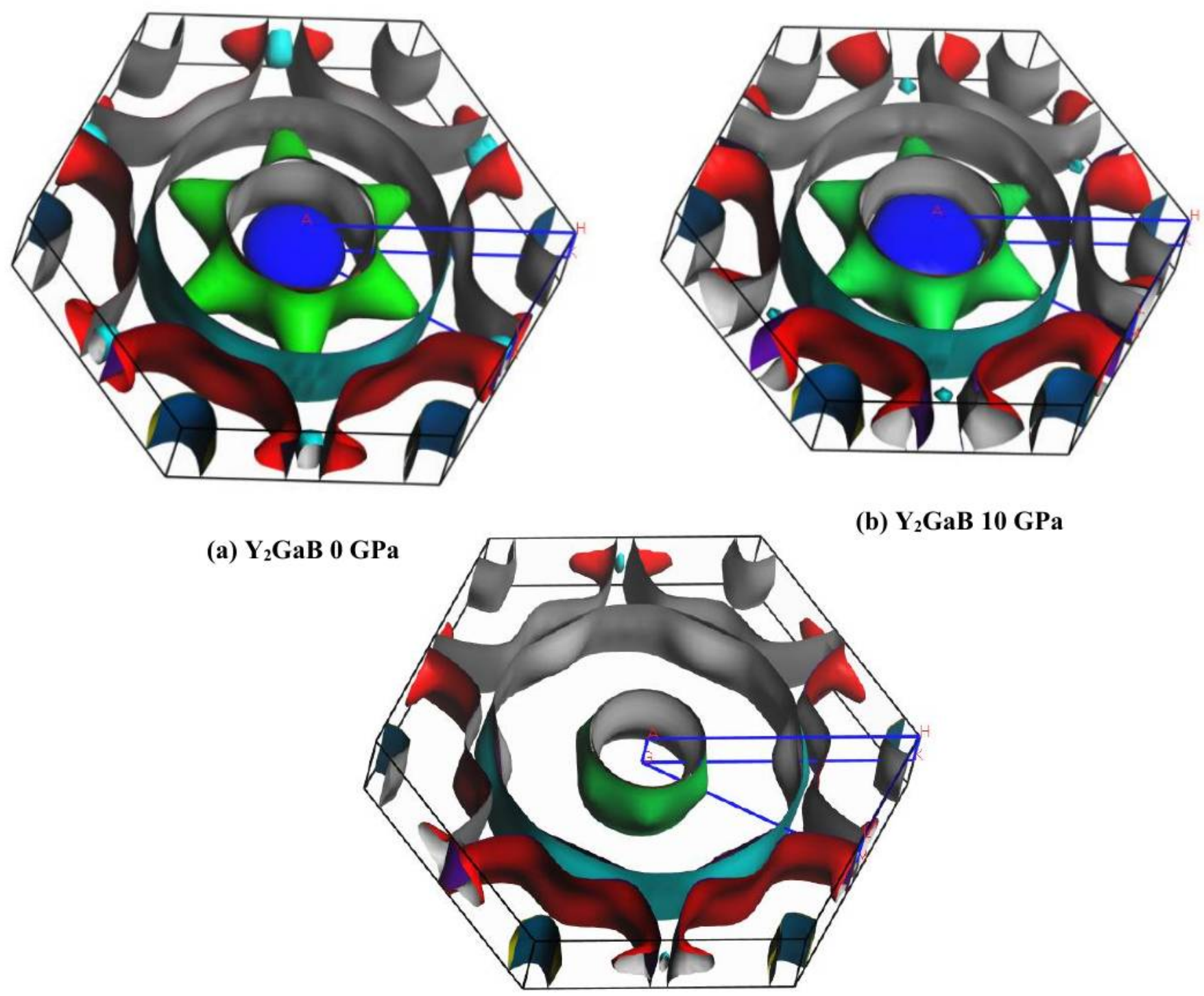

(c) $Y_2InB$ 0 GPa

**Figure 16** Combined Fermi surfaces of (a) $Y_2GaB$ at 0 GPa, (b) $Y_2GaB$ at 10 GPa, and (c) $Y_2InB$ at 0 GPa, obtained by superimposing the individual sheets within the first Brillouin zone.

Replacing Ga with In alters the central region more distinctly. Ambient-pressure $Y_2InB$ lacks the innermost compact electron pocket of $Y_2GaB$ and instead begins with smaller ring-like sheets. Its outer Fermi surface still contains broad toroidal components, sixfold-corrugated sheets, networks connected to the zone boundary, and small edge pockets. The substitution thus lowers the number of resolved sheets while preserving the coexistence of closed and boundary-connected components. This difference agrees with the near-$E_F$ band arrangement in Section 3.5. The present surfaces do not carry sheet-resolved orbital information, so the DOS supports only a collective, not sheet-specific, orbital interpretation.

The combined views (**Figure 16**) reveal a three-dimensional, anisotropic topology rather than a strictly two-dimensional one. Broad basal-plane rings and sixfold networks coexist with finite axial extension and compact pockets, as also reported for nanolamellar carbides **[70–72]**. These geometries indicate reciprocal-space anisotropy, but they do not determine carrier concentration, electron or hole dominance, mobility, conductivity, effective mass, or nesting. The colors in **Figure 16** distinguish the plotted sheets only.

### 3.8 Electron-Density Difference and Bonding Characteristics

To examine how pressure and Ga/In substitution redistribute the valence charge, the electron-density difference was calculated as $\Delta\rho(r) = \rho_{crystal}(r) - \sum_i \rho_i^{free}(r)$. The reference density is the sum of neutral free-atom densities placed at the crystal sites. **Figure 17** shows the (110) sections for $Y_2GaB$ at 0 and 10 GPa and $Y_2InB$ at 0 GPa. Red indicates accumulation and blue indicates depletion of electronic charge. Since each panel has its own contour interval, the maps can be compared by the location and shape of the redistributed density, not only by color intensity.

The ambient-pressure maps share a clear feature. Electron density is depleted around Y and accumulates around B. The accumulated charge follows the Y-B contacts, revealing a directional component in these bonds, whereas the changes around Ga and In are weaker and more diffuse suggesting metallic bonding. In $Y_2InB$, the redistributed density near the In layer spreads over a somewhat broader region of the section. This is compatible with the larger axial spacing and the In-5*p* contribution seen in the DOS, although the different contour scales rule out a quantitative comparison of charge transfer. The $Y_2GaB$ map at 10 GPa keeps the same basic pattern. Pressure mainly reshapes the interstitial density: accumulation around B becomes more confined, while depletion extends through more of the neighboring Y- and Ga-containing regions. A comparable bonding hierarchy, strong transition-metal-B interactions separated by a more weakly bonded A-containing layer, is common in layered MAX and MAB phases **[75–79]**.

The charge analyses support the above picture. Mulliken and Hirshfeld partitioning both give positive Y, negative B, and nearly neutral Ga or In, even though the absolute values differ between the two schemes. For $Y_2GaB$ at 0 GPa, the Mulliken/Hirshfeld charges are +0.39/+0.08 *e* on Y and −0.72/−0.26 *e* on B. For $Y_2InB$, the corresponding B charges are −0.74/−0.28 *e*. Compression changes the magnitudes only slightly and leaves the Y-to-B transfer direction unchanged. These values are population charges, not formal oxidation states, and their size depends on the chosen partitioning and orbital projection **[34–36]**.

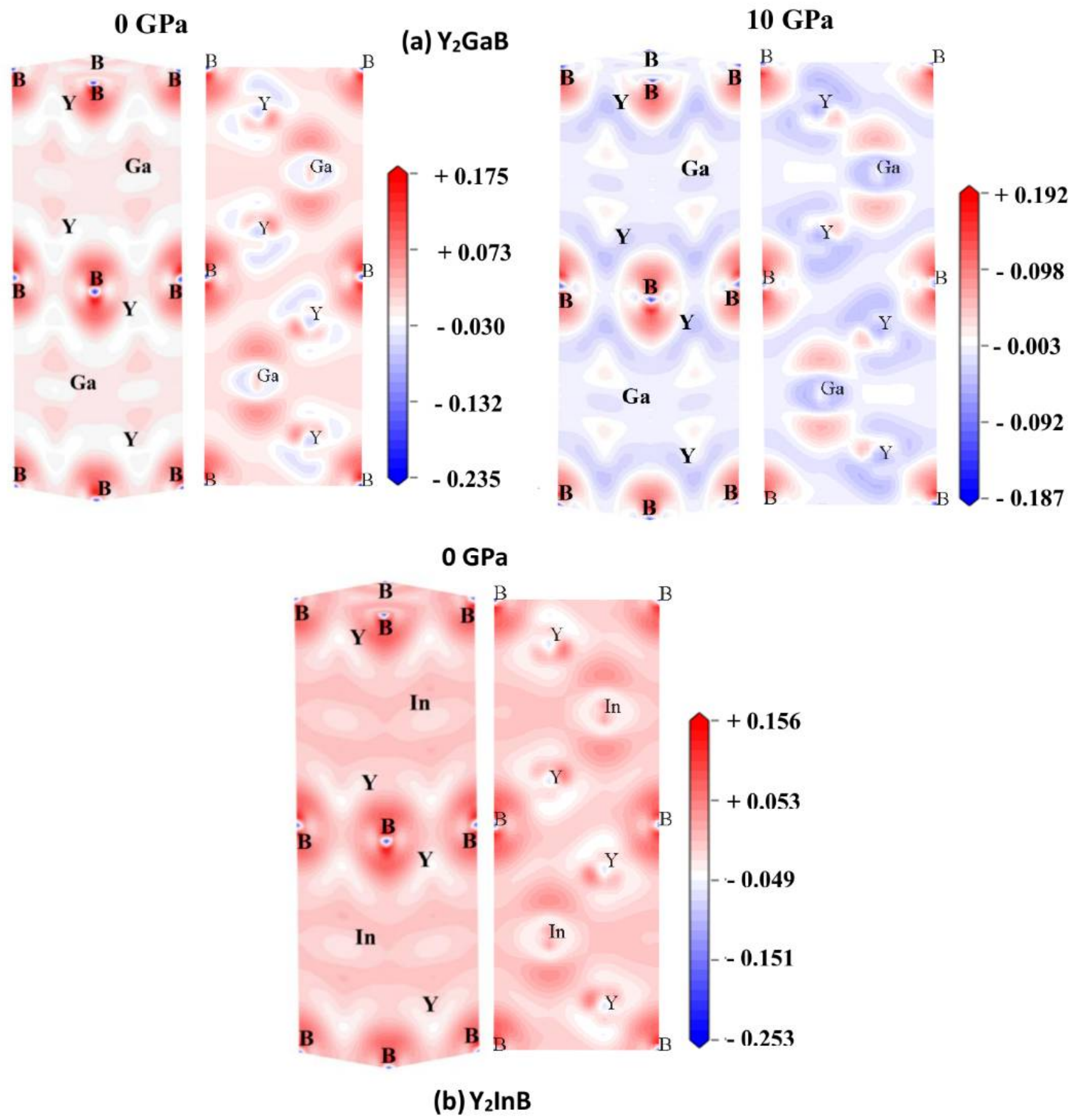

**Figure 17** Electron-density-difference maps in the (110) plane for (a) $Y_2GaB$ at 0 and 10 GPa and (b) $Y_2InB$ at 0 GPa. Red and blue represent electron accumulation and depletion, respectively. The contour intervals are −0.235 to +0.175, −0.187 to +0.192, and −0.253 to +0.156 e $Å^{-3}$ for $Y_2GaB$ at 0 GPa, $Y_2GaB$ at 10 GPa, and $Y_2InB$ at 0 GPa, respectively.

The Y-B Mulliken overlap population is also positive. At 0 GPa, $P_\mu$ is 1.14 at 2.63 Å in $Y_2GaB$ and 1.17 at 2.62 Å in $Y_2InB$. For $Y_2GaB$ at 10 GPa, the Y-B distance shortens to 2.56 Å, but $P_\mu$ decreases to 1.07. Bond contraction therefore does not produce a proportional increase in the projected covalent population. Together with the Y-4*d*/B-2*p* spectral overlap, these results identify Y-B bonds with both ionic and covalent contributions. The finite DOS and multiband Fermi surfaces show that this local bonding exists within a metallic electronic background. The same combination has been reported for other layered ternary borides and MAX/MAB phases **[15,75–79]**. Its precise numerical description, however, remains dependent on the selected population scheme.

## 3.9 Optical Properties

The optical response was examined for the elastically stable states, $Y_2GaB$ at 0, 5, and 10 GPa and $Y_2InB$ at 0 GPa, with electric field polarization along [100] and [001] directions (**Figure 18**). In each case, the Drude intraband term overlaps with low-energy interband transitions and produces a strong response at the lower end of the spectrum. This behavior agrees with the multiple bands crossing $E_F$ and the finite Fermi-level DOS. Absorption and optical conductivity therefore begin at the lowest sampled energies, with no semiconductor-like threshold. The real dielectric function $\varepsilon_1(\omega)$ is strongly negative near the lower boundary, crosses zero several times, and becomes weakly positive at higher energy. The main structures in $\varepsilon_2(\omega)$, $n(\omega)$, and $k(\omega)$ lie mostly below about 6 eV. Because Drude damping and spectral broadening control the low-frequency metallic limit, the first plotted $n$ and $k$ values are finite-frequency results rather than static constants. Related MAX and MAB phases also show pronounced polarization dependence in optical conductivity and dielectric loss **[37,38,80–82]**.

Pressure reorganizes the $Y_2GaB$ spectra rather than shifting every feature by the same amount. The highest-energy zero crossing of $\varepsilon_1(\omega)$ moves from 12.70 and 12.73 eV at 0 GPa to 13.80 and 13.72 eV at 10 GPa for [100] and [001] polarizations, respectively. In $\varepsilon_2(\omega)$, the prominent axial feature shifts from about 4.17 to 4.77 eV and loses some intensity. The principal [100] absorption maximum moves from 7.45 to 8.54 eV. For [001], the dominant maximum first shifts from 9.29 eV at 0 GPa to 9.82 eV at 5 GPa, after which spectral weight is transferred to a stronger feature near 7.58 eV at 10 GPa. The two polarizations thus respond differently to compression, both in peak position and relative intensity. Similar pressure- and composition-dependent rearrangements have been reported for metallic layered carbides and related systems **[83,84]**.

At the lowest photon energy, the reflectivity is approximately 0.981 for every state and polarization, as expected from the Drude-dominated metallic response. Away from this low-energy limit, the main $Y_2GaB$ reflectivity maximum for [100] shifts from 7.84 eV at 0 GPa to 9.07 eV at 10 GPa, while the corresponding [001] feature moves from 10.20 to 11.59 eV. Their amplitudes do not vary monotonically. The conductivity maxima follow a comparable trend: the [100] peak moves from 6.76 to 7.78 eV, and the stronger [001] peak from 4.43 to 5.07 eV. A clearer signature of collective excitation appears in the loss function. Its dominant maxima occur at 13.02 and 12.69 eV for ambient-pressure $Y_2GaB$ and shift to 14.53 and 13.70 eV at 10 GPa for [100] and [001] polarizations. Their proximity to the high-energy zero crossings of $\varepsilon_1(\omega)$, together with the small $\varepsilon_2(\omega)$ values in the same region, supports a plasma-like loss process. These energies are characteristic loss maxima, not independently fitted or experimentally measured free-electron plasma frequencies **[81,82]**.

$Y_2InB$ shows the same metallic response at low photon energy but differs more clearly at higher energies. Its principal absorption maxima lie at 7.42 eV for [100] and 9.54 eV for [001] polarization, close to the corresponding ambient-pressure $Y_2GaB$ features. The main finite-energy reflectivity maxima reach 0.513 at 7.99 eV and 0.677 at 10.70 eV, respectively. Conductivity peaks occur at 6.67 and 4.15 eV, while the loss-function maxima appear at 13.16 and 13.29 eV, respectively. The axial loss peak reaches 5.770 on the plotted dimensionless scale and gives the strongest polarization contrast among the ambient-pressure loss spectra.

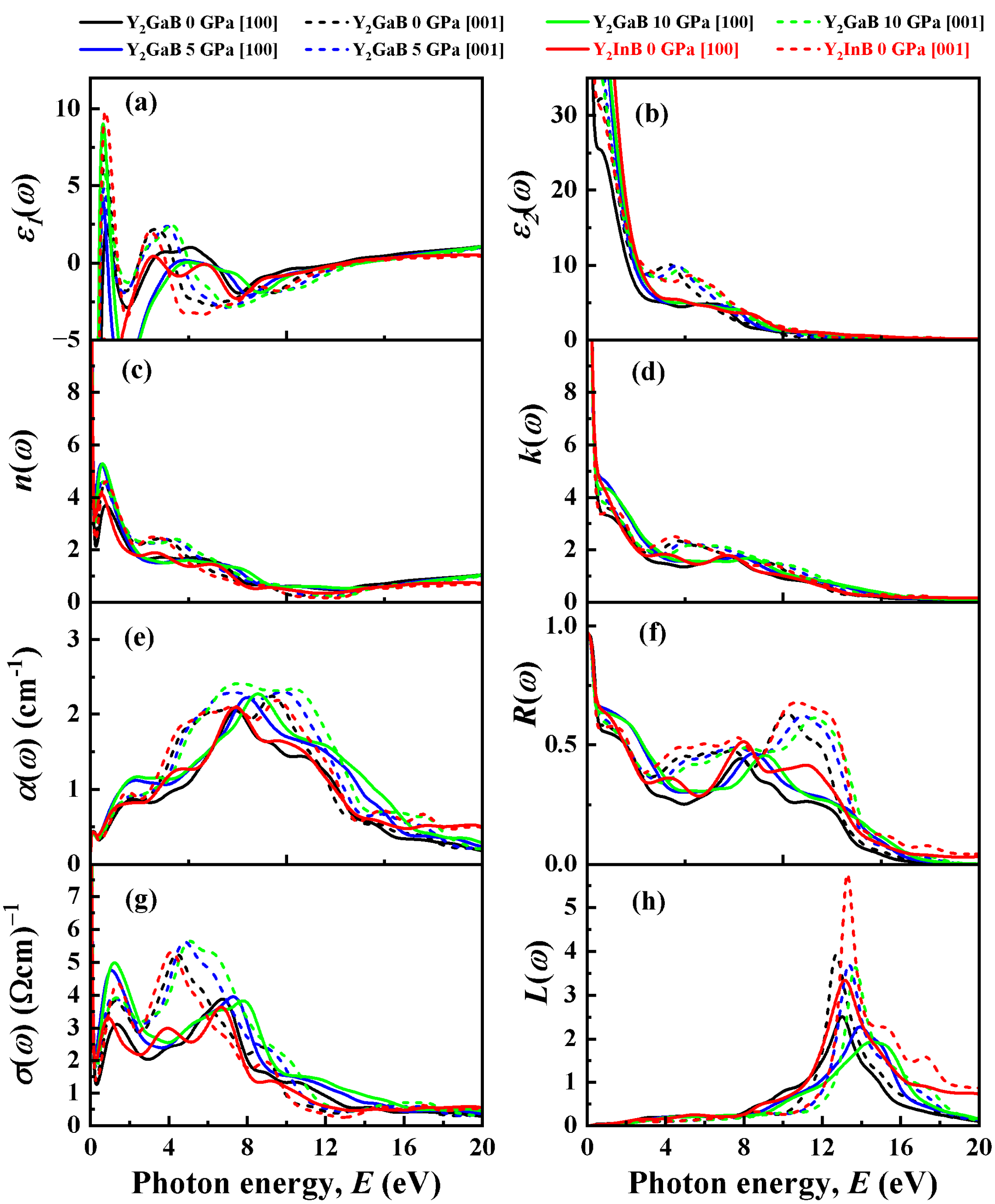

**Figure 18** Polarization-resolved optical properties of $Y_2GaB$ at 0, 5, and 10 GPa and $Y_2InB$ at 0 GPa along [100] (solid) and [001] (dashed): (a) $\varepsilon_1(\omega)$, (b) $\varepsilon_2(\omega)$, (c) $n(\omega)$, (d) $k(\omega)$, (e) $\alpha(\omega)$, (f) $R(\omega)$, (g) $\sigma(\omega)$, and (h) $L(\omega)$. Black, blue, green, and red denote $Y_2GaB$ at 0, 5, and 10 GPa and $Y_2InB$ at 0 GPa, respectively. $E$ is in eV, $\alpha$ in $cm^{-1}$, and $\sigma$ in $\Omega^{-1}$ $cm^{-1}$; the remaining quantities are dimensionless.

These differences are consistent with changes in A-site *p*-state energies, the arrangement of bands near $E_F$, lattice geometry, and orbital mixing rather than atomic mass alone **[80,84]**. Taken together, the spectra show that both compounds remain metallic and optically anisotropic. Compression generally raises the characteristic energies of the principal $Y_2GaB$ reflectivity, conductivity, and loss features, whereas the absorption response is not monotonic.

The calculated optical parameters demonstrate that both the state of polarization and pressure can be used to tune the optical properties of $Y_2GaB$. Both the MAB compounds are efficient reflector of light in the visible region and they are good absorbers of ultraviolet light.

## 3.10 Phonon Dynamics and Dynamical Stability

The vibrational response was examined from the unit-cell phonon dispersions and total phonon densities of states (PhDOS) of $Y_2GaB$ and $Y_2InB$ at 0 and 10 GPa (Figure 20). Finite displacements were applied **[39,85,86]** in CASTEP to the eight-atom crystallographic cell. The calculations used an 8 × 8 × 2 electronic *k*-point mesh, plane wave cutoffs of 408.2 eV for $Y_2GaB$ and 390.0 eV for $Y_2InB$, and a smearing width of 0.01 eV. In a harmonic calculation, an extended imaginary branch indicates an unstable vibrational eigenmode, whereas a minute residual close to Γ may arise from imperfect translational invariance or numerical convergence **[87,88]**. No substantial imaginary branch is resolved in any of the four plotted spectra. The results therefore support dynamical stability at the sampled *q* points within this restricted unit-cell treatment.

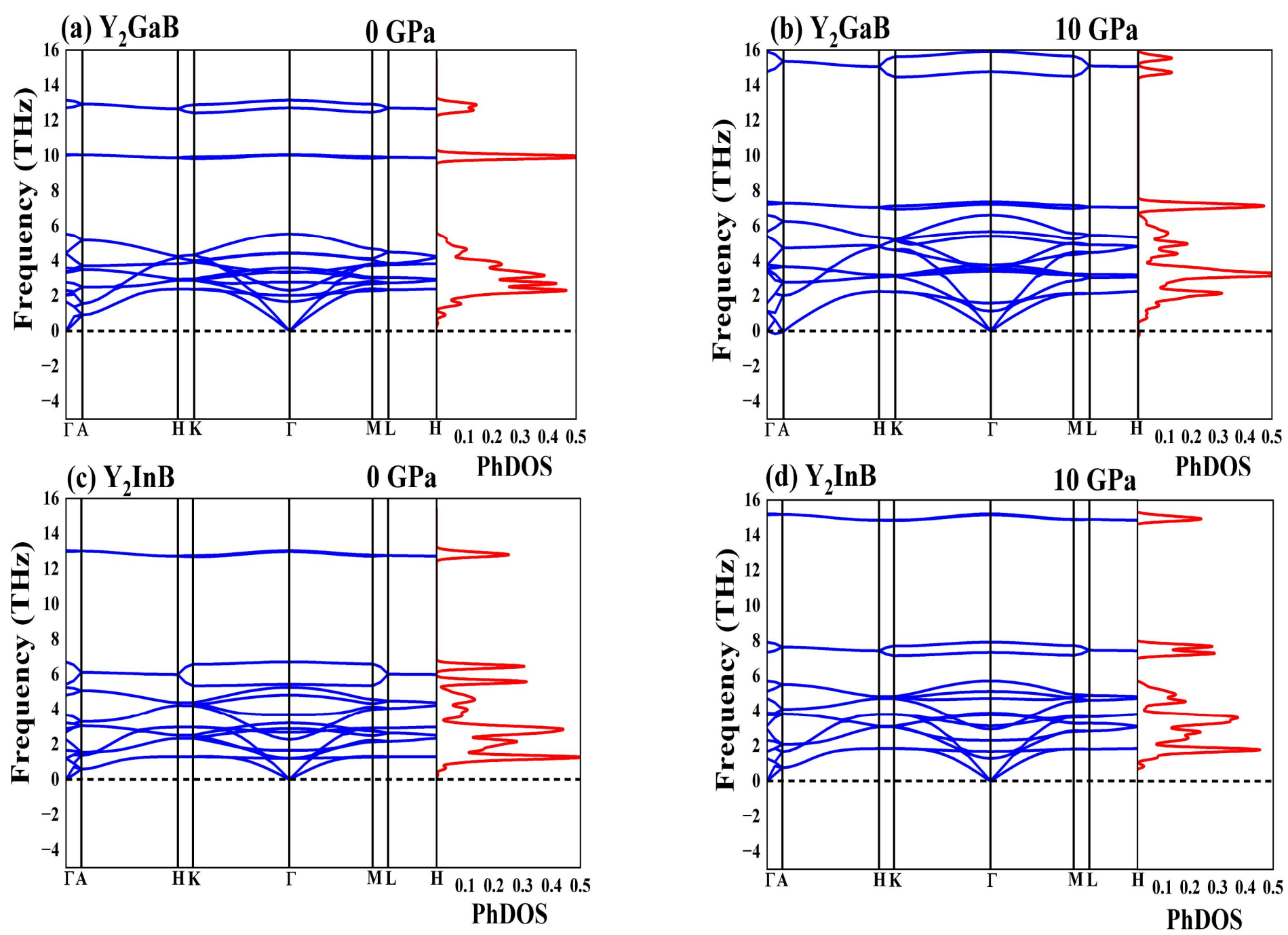

**Figure 19** Unit-cell phonon dispersions and corresponding total PhDOS of (a) $Y_2GaB$ at 0 GPa, (b) $Y_2GaB$ at 10 GPa, (c) $Y_2InB$ at 0 GPa, and (d) $Y_2InB$ at 10 GPa along Γ–A–H–K–Γ–M–L–H. Frequencies are given in THz, and the total PhDOS is reported in arbitrary units.

At ambient pressure, the three acoustic branches of $Y_2GaB$ approach zero smoothly at Γ, while the optical modes remain at positive frequency along the sampled path. Most acoustic and lower optical states lie below about 6 THz, followed by comparatively flat upper groups near 10 and 13 THz. Compression to 10 GPa hardens several branches and pushes the upper edge to approximately 15-16 THz. The accompanying PhDOS redistribution is nonuniform, so pressure does more than shift the spectrum rigidly. Even so, no negative-frequency region develops in the unit-cell result. $Y_2GaB$ is therefore both elastically stable and dynamically stable within this calculation at 0 and 10 GPa.

$Y_2InB$ follows a broadly similar vibrational evolution. At 0 GPa, the acoustic modes merge into a closely spaced low-frequency manifold, and the highest optical branches reach roughly 13 THz. Several modes harden at 10 GPa, raising the upper limit to about 15-16 THz, although individual branches respond by different amounts. The total PhDOS retains separate low, intermediate and high-frequency features. Its lower ambient-pressure bandwidth relative to $Y_2GaB$ is more reasonably attributed to the combined effects of atomic masses, lattice geometry, and interatomic force constants than to the In mass alone. Comparable composition-dependent phonon manifolds have been reported for layered MAX and MAB compounds **[89–92]**. The positive unit-cell spectrum at 10 GPa does not supersede the elastic result: $Y_2InB$ is elastically stable only at 0 GPa and remains elastically unstable at 10 GPa.

The dynamical-stability statement must remain tied to the numerical representation used for **Figure 19**. Tests with expanded supercells produced imaginary modes, revealing a sensitivity of the low-frequency spectrum to cell size and force constant treatment **[85–88]**. The unit-cell dispersions are therefore positive at 0 and 10 GPa, but a method-independent stability assignment requires systematic convergence with respect to supercell size and the real-space force constant range. This qualification concerns the phonon calculation only and does not alter the pressure-corrected elastic-stability conclusions.

## 4 Conclusions

$Y_2GaB$ and $Y_2InB$ were examined in the layered hexagonal $P6_3/mmc$ structure to establish their pressure-dependent structure-property relationships. Both lattices contract anisotropically under pressure, with the basal plane responding more strongly than the axial direction. According to the pressure-corrected elastic stability criteria, $Y_2GaB$ is elastically stable at 0, 5, and 10 GPa but elastically unstable at 15 and 20 GPa, whereas $Y_2InB$ is elastically stable only at 0 GPa. Within the elastically stable $Y_2GaB$ pressure range, compression increases the elastic stiffness, sound velocities, Debye temperature, empirical melting-temperature estimate, and calculated thermal-transport indicators. Its Slack-estimated lattice thermal conductivity rises from 0.605 W $m^{-1}$ $K^{-1}$ at 0 GPa to 1.543 W $m^{-1}$ $K^{-1}$ at 10 GPa. Extremely low minimum thermal conductivities of the two compounds suggest their potential as thermal barrier coating material at low temperature – limited by the fairly low melting temperatures.

The calculations reveal multiband metallic electronic structures dominated near the Fermi level by Y-4*d* states, with secondary B-2*p* and Ga-4*p* or In-5*p* contributions. Spin-orbit coupling locally modifies selected bands without opening a gap. $Y_2GaB$ retains nine band-resolved Fermi-surface sheets at 0 and 10 GPa, while ambient-pressure $Y_2InB$ has eight; pressure reshapes the $Y_2GaB$ surfaces without verified sheet creation, disappearance, or connectivity change, so no Lifshitz transition is established. The Fermi surface segments exhibit both electron and hole characters.

The combined electron-density difference, charge-partitioning, overlap-population, and DOS evidence supports mixed ionic-covalent Y-B interactions with complementary delocalized metallic character.

Both compounds show a metallic Drude response and clear [100]/[001] optical anisotropy. The compounds are efficient reflectors of visible radiation and good absorber of ultraviolet light. Their unit-cell phonon dispersions contain positive frequencies at 0 and 10 GPa, indicating that both are dynamically stable within the adopted unit-cell calculations. $Y_2InB$ at 10 GPa remains elastically unstable despite its positive unit-cell phonon spectrum, confirming that elastic and dynamical stability describe different perturbations and must not be conflated.

$Y_2GaB$ therefore has the broader elastic-stability window and exhibits systematic pressure-driven modification of its mechanical, thermophysical, electronic, Fermi-surface, and optical responses within that range. $Y_2InB$ has a narrower elastic-stability window, although it retains metallic and optically anisotropic characteristics together with dynamically stable unit-cell phonon spectra at the sampled pressures. Expanded-supercell calculations produced imaginary modes, indicating sensitivity of the low-frequency phonon response to cell size and force-constant treatment. The study thus establishes a comparative structure-property framework, identifies the mechanically viable pressure ranges, and defines the numerical limitations relevant to future experimental validation and more systematic phonon-convergence analysis.

## Data availability

The data sets generated and/or analyzed in this study are available from the corresponding author on reasonable request.

## Declaration of interest

The authors declare no competing interests.

## CRediT authorship contribution statement

**Md. Rabi Us Sani Shapon:**Software, Methodology, Formal analysis, Data curation, Visualization, Writing-original draft, review & editing; **Md. Asif Afzal:** Software, Formal analysis, Data curation, Visualization, Writing-draft, review & editing; **Jubair Hossan Abir:** Formal analysis, Visualization, Writing- draft, review & editing; **Raihana Shams Islam:** Conceptualization, Supervision, Validation, Administration, Writing- Reviewing and Editing; **Saleh Hasan Naqib:** Conceptualization, Supervision, Validation, Administration, Writing-Reviewing and Editing.

## Declaration of Generative AI and AI-assisted technologies in the writing process

During the preparation of this work, the authors used ChatGPT (OpenAI) in order to improve the language, grammar, readability, and organization of the manuscript. After using this tool, the authors reviewed and edited the content as needed and take full responsibility for the content of the published article.

## References

[1] M.W. Barsoum, The $M_{n+1}AX_n$ phases: A new class of solids; thermodynamically stable nanolaminates, *Progress in Solid State Chemistry*, 28 (2000) 201–281. https://doi.org/10.1016/S0079-6786(00)00006-6.

[2] Z.M. Sun, Progress in research and development on MAX phases: a family of layered ternary compounds, *International Materials Reviews*, 56 (2011) 143–166. https://doi.org/10.1179/1743280410Y.0000000001.

[3] M. Sokol, V. Natu, S. Kota, M.W. Barsoum, On the chemical diversity of the MAX phases, *Trends in Chemistry*, 1 (2019) 210–223. https://doi.org/10.1016/j.trechm.2019.02.016.

[4] M.W. Barsoum, M. Radovic, Elastic and mechanical properties of the MAX phases, *Annual Review of Materials Research*, 41 (2011) 195–227. https://doi.org/10.1146/annurev-matsci-062910-100448.

[5] M.W. Barsoum, T. El-Raghy, Synthesis and characterization of a remarkable ceramic: $Ti_3SiC_2$, *Journal of the American Ceramic Society*, 79 (1996) 1953–1956. https://doi.org/10.1111/j.1151-2916.1996.tb08018.x.

[6] M. Dahlqvist, B. Alling, J. Rosén, Stability trends of MAX phases from first principles, *Physical Review B*, 81 (2010) 220102. https://doi.org/10.1103/PhysRevB.81.220102.

[7] M. Khazaei, M. Arai, T. Sasaki, M. Estili, Y. Sakka, Trends in electronic structures and structural properties of MAX phases: a first-principles study on $M_2AlC$ (M = Sc, Ti, Cr, Zr, Nb, Mo, Hf, or Ta), $M_2AlN$, and hypothetical $M_2AlB$ phases, *Journal of Physics: Condensed Matter*, 26 (2014) 505503. https://doi.org/10.1088/0953-8984/26/50/505503.

[8] M. Ade, H. Hillebrecht, Ternary borides $Cr_2AlB_2$, $Cr_3AlB_4$, and $Cr_4AlB_6$: the first members of the series $(CrB_2)_nCrAl$ with n = 1, 2, 3 and a unifying concept for ternary borides as MAB-phases, *Inorganic Chemistry*, 54 (2015) 6122–6135. https://doi.org/10.1021/acs.inorgchem.5b00049.

[9] L.T. Alameda, P. Moradifar, Z.P. Metzger, N. Alem, R.E. Schaak, Topochemical deintercalation of Al from MoAlB: Stepwise etching pathway, layered intergrowth structures, and two-dimensional MBene, *Journal of the American Chemical Society*, 140 (2018) 8833–8840. https://doi.org/10.1021/jacs.8b04705.

[10] T. Rackl, L. Eisenburger, R. Niklaus, D. Johrendt, Syntheses and physical properties of the MAX phase boride $Nb_2SB$ and the solid solutions $Nb_2SB_xC_{1-x}$ (x = 0-1), *Physical Review Materials*, 3 (2019) 054001. https://doi.org/10.1103/PhysRevMaterials.3.054001.

[11] T. Rackl, D. Johrendt, The MAX phase borides $Zr_2SB$ and $Hf_2SB$, *Solid State Sciences*, 106 (2020) 106316. https://doi.org/10.1016/j.solidstatesciences.2020.106316.

[12] J. Mou, S. Li, W. Zhang, W. Xu, S. Fan, G. Bei, Deintercalation of Al from MoAlB by molten salt etching to achieve a $Mo_2AlB_2$ compound and 2D MoB nanosheets, *Journal of Advanced Ceramics*, 12 (2023) 943–953. https://doi.org/10.26599/JAC.2023.9220729.

[13] G. Surucu, Investigation of structural, electronic, anisotropic elastic, and lattice dynamical properties of MAX phases borides: an ab-initio study on hypothetical $M_2AB$ (M = Ti, Zr, Hf; A = Al, Ga, In) compounds, *Materials Chemistry and Physics*, 203 (2018) 106–117. https://doi.org/10.1016/j.matchemphys.2017.09.050.

[14] M.A. Ali, M.M. Hossain, M.M. Uddin, M.A. Hossain, A.K.M.A. Islam, S.H. Naqib, Physical properties of new MAX phase borides $M_2SB$ (M = Zr, Hf and Nb) in comparison with conventional MAX phase carbides $M_2SC$ (M = Zr, Hf and Nb): comprehensive insights, *Journal of Materials Research and Technology*, 11 (2021) 1000–1018. https://doi.org/10.1016/j.jmrt.2021.01.068.

[15] Y. Bai, X. Qi, X. He, D. Sun, F. Kong, Y. Zheng, R. Wang, A.I. Duff, Phase stability and weak metallic bonding within ternary-layered borides CrAlB, $Cr_2AlB_2$, $Cr_3AlB_4$, and $Cr_4AlB_6$,

*Journal of the American Ceramic Society*, 102 (2019) 3715–3727. https://doi.org/10.1111/jace.16206.
[16] S.H. Sakib, M.N. Uddin, A.K.M.N. Ishtiaq, S. Islam, M.R. Rana, N.A. Roktim, M.E. Hossain, M.A. Ali, K. Hoque, Prediction and stability of $Y_2AB$ (A = Ga, In) MAX phase borides via DFT study, *Discover Materials*, 6 (2026) 54. https://doi.org/10.1007/s43939-025-00459-5.
[17] A. Bouhemadou, R. Khenata, Prediction study of structural and elastic properties under the pressure effect of $M_2GaC$ (M = Ti, V, Nb, Ta), *Journal of Applied Physics*, 102 (2007) 043528. https://doi.org/10.1063/1.2773634.
[18] M.W. Qureshi, X. Ma, G. Tang, R. Paudel, Ab initio predictions of structure and physical properties of the $Zr_2GaC$ and $Hf_2GaC$ MAX phases under pressure, *Scientific Reports*, 11 (2021) 3260. https://doi.org/10.1038/s41598-021-82402-1.
[19] S. Chen, H. Ma, X. Zhang, L. Chen, H. Wu, H. Li, First-principles calculations of $M_2SB$ (M = Hf, Zr and Nb) under high pressure, *Physics Letters A*, 547 (2025) 130551. https://doi.org/10.1016/j.physleta.2025.130551.
[20] S.I. Ranganathan, M. Ostoja-Starzewski, Universal elastic anisotropy index, *Physical Review Letters*, 101 (2008) 055504. https://doi.org/10.1103/PhysRevLett.101.055504.
[21] F. Mouhat, F.-X. Coudert, Necessary and sufficient elastic stability conditions in various crystal systems, *Physical Review B*, 90 (2014) 224104. https://doi.org/10.1103/PhysRevB.90.224104.
[22] S. Baroni, S. de Gironcoli, A. Dal Corso, P. Giannozzi, Phonons and related crystal properties from density-functional perturbation theory, *Reviews of Modern Physics*, 73 (2001) 515–562. https://doi.org/10.1103/RevModPhys.73.515.
[23] P. Hohenberg, W. Kohn, Inhomogeneous electron gas, *Physical Review*, 136 (1964) B864–B871. https://doi.org/10.1103/PhysRev.136.B864.
[24] W. Kohn, L.J. Sham, Self-consistent equations including exchange and correlation effects, *Physical Review*, 140 (1965) A1133–A1138. https://doi.org/10.1103/PhysRev.140.A1133.
[25] M.D. Segall, P.J.D. Lindan, M.J. Probert, C.J. Pickard, P.J. Hasnip, S.J. Clark, M.C. Payne, First-principles simulation: ideas, illustrations and the CASTEP code, *Journal of Physics: Condensed Matter*, 14 (2002) 2717–2744. https://doi.org/10.1088/0953-8984/14/11/301.
[26] J.P. Perdew, K. Burke, M. Ernzerhof, Generalized gradient approximation made simple, *Physical Review Letters*, 77 (1996) 3865–3868. https://doi.org/10.1103/PhysRevLett.77.3865.
[27] D. Vanderbilt, Soft self-consistent pseudopotentials in a generalized eigenvalue formalism, *Physical Review B*, 41 (1990) 7892–7895. https://doi.org/10.1103/PhysRevB.41.7892.
[28] D.D. Koelling, B.N. Harmon, A technique for relativistic spin-polarised calculations, *Journal of Physics C: Solid State Physics*, 10 (1977) 3107–3114. https://doi.org/10.1088/0022-3719/10/16/019.
[29] H.J. Monkhorst, J.D. Pack, Special points for Brillouin-zone integrations, *Physical Review B*, 13 (1976) 5188–5192. https://doi.org/10.1103/PhysRevB.13.5188.
[30] B.G. Pfrommer, M. Côté, S.G. Louie, M.L. Cohen, Relaxation of crystals with the quasi-Newton method, *Journal of Computational Physics*, 131 (1997) 233–240. https://doi.org/10.1006/jcph.1996.5612.
[31] O.H. Nielsen, R.M. Martin, First-principles calculation of stress, *Physical Review Letters*, 50 (1983) 697–700. https://doi.org/10.1103/PhysRevLett.50.697.
[32] P. Giannozzi et al., QUANTUM ESPRESSO: a modular and open-source software project for quantum simulations of materials, *Journal of Physics: Condensed Matter*, 21 (2009) 395502. https://doi.org/10.1088/0953-8984/21/39/395502.
[33] P.E. Blöchl, Projector augmented-wave method, *Physical Review B*, 50 (1994) 17953–17979. https://doi.org/10.1103/PhysRevB.50.17953.

[34] R.S. Mulliken, Electronic population analysis on LCAO–MO molecular wave functions. II. Overlap populations, bond orders, and covalent bond energies, *Journal of Chemical Physics*, 23 (1955) 1841–1846. https://doi.org/10.1063/1.1740589.
[35] F.L. Hirshfeld, Bonded-atom fragments for describing molecular charge densities, *Theoretica Chimica Acta*, 44 (1977) 129–138. https://doi.org/10.1007/BF00549096.
[36] M.D. Segall, R. Shah, C.J. Pickard, M.C. Payne, Population analysis of plane-wave electronic structure calculations of bulk materials, *Physical Review B*, 54 (1996) 16317–16320. https://doi.org/10.1103/PhysRevB.54.16317.
[37] M. Gajdoš, K. Hummer, G. Kresse, J. Furthmüller, F. Bechstedt, Linear optical properties in the projector-augmented wave methodology, *Physical Review B*, 73 (2006) 045112. https://doi.org/10.1103/PhysRevB.73.045112.
[38] P. Drude, Zur Elektronentheorie der Metalle, *Annalen der Physik*, 306 (1900) 566–613. https://doi.org/10.1002/andp.19003060312.
[39] K. Parlinski, Z.Q. Li, Y. Kawazoe, First-principles determination of the soft mode in cubic $ZrO_2$, *Physical Review Letters*, 78 (1997) 4063–4066. https://doi.org/10.1103/PhysRevLett.78.4063.
[40] M.R. Rana, S. Islam, K. Hoque, G.G. Biswas, M.E. Hossain, S.H. Naqib, M.A. Ali, DFT prediction of the stability and physical properties of $M_2GaB$ (M = Sc, V, Nb, Ta), *Journal of Materials Research and Technology*, 24 (2023) 7795–7815. https://doi.org/10.1016/j.jmrt.2023.05.008.
[41] S. Islam, M.R. Rana, K. Hoque, G.G. Biswas, M.E. Hossain, M.M. Hossain, M.M. Uddin, S.H. Naqib, M.A. Ali, A comprehensive exploration of the physical properties of $M_2GaB$ (M = Ti, Zr, Mo, Hf) through DFT method, *Results in Materials*, 19 (2023) 100438. https://doi.org/10.1016/j.rinma.2023.100438.
[42] M.A. Ali, M.M. Hossain, M.M. Uddin, A.K.M.A. Islam, S.H. Naqib, Understanding the improvement of thermo-mechanical and optical properties of 212 MAX phase borides $Zr_2AB_2$ (A = In, Tl), *Journal of Materials Research and Technology*, 15 (2021) 2227–2241. https://doi.org/10.1016/j.jmrt.2021.09.042.
[43] M.A. Rahman, M.I. Kholil, R. Khatun, S. Sarker, M. Hasan, M.A. Ali, M.Z. Rahaman, K.M. Hossain, M.I. Haque, M.Z. Hasan, High pressure study of new type of MAX phases: $Hf_2AB_2$ (A = In, Sn), *physica status solidi (b)*, 260 (2023) 2200102. https://doi.org/10.1002/pssb.202200102.
[44] T.H.K. Barron, M.L. Klein, Second-order elastic constants of a solid under stress, *Proceedings of the Physical Society*, 85 (1965) 523–532. https://doi.org/10.1088/0370-1328/85/3/313.
[45] G. Surucu, A. Gencer, X. Wang, O. Surucu, Lattice dynamical and thermo-elastic properties of $M_2AlB$ (M = V, Nb, Ta) MAX phase borides, *Journal of Alloys and Compounds*, 819 (2020) 153256. https://doi.org/10.1016/j.jallcom.2019.153256.
[46] M.A. Hadi, M.Z. Alam, I. Ahmed, A.M.M.T. Karim, S.H. Naqib, A. Chroneos, A.K.M.A. Islam, A density functional theory approach to the effects of C and N substitution at the B-site of the first boride MAX phase $Nb_2SB$, *Materials Today Communications*, 29 (2021) 102910. https://doi.org/10.1016/j.mtcomm.2021.102910.
[47] R. Hill, The elastic behaviour of a crystalline aggregate, *Proceedings of the Physical Society. Section A*, 65 (1952) 349–354. https://doi.org/10.1088/0370-1298/65/5/307.
[48] Y.X. Wang, Z.X. Yan, W. Liu, G.L. Zhou, Structure stability, mechanical properties and thermal conductivity of the new hexagonal ternary phase $Ti_2InB_2$ under pressure, *Philosophical Magazine*, 100 (2020) 2054–2067. https://doi.org/10.1080/14786435.2020.1754485.
[49] M.A. Ali, M.M. Hossain, M.M. Uddin, A.K.M.A. Islam, S.H. Naqib, The rise of 212 MAX phase borides: DFT insights into the physical properties of $Ti_2PB_2$, $Zr_2PbB_2$, and $Nb_2AB_2$ (A =

P, S) for thermomechanical applications, *ACS Omega*, 8 (2023) 954–968. https://doi.org/10.1021/acsomega.2c06331.
[50] S.F. Pugh, XCII. Relations between the elastic moduli and the plastic properties of polycrystalline pure metals, *The London, Edinburgh, and Dublin Philosophical Magazine and Journal of Science*, 45 (1954) 823–843. https://doi.org/10.1080/14786440808520496.
[51] D.G. Pettifor, Theoretical predictions of structure and related properties of intermetallics, *Materials Science and Technology*, 8 (1992) 345–349. https://doi.org/10.1179/mst.1992.8.4.345.
[52] Z. Sun, D. Music, R. Ahuja, J.M. Schneider, Theoretical investigation of the bonding and elastic properties of nanolayered ternary nitrides, *Physical Review B*, 71 (2005) 193402. https://doi.org/10.1103/PhysRevB.71.193402.
[53] Y. Tian, B. Xu, Z. Zhao, Microscopic theory of hardness and design of novel superhard crystals, *International Journal of Refractory Metals and Hard Materials*, 33 (2012) 93–106. https://doi.org/10.1016/j.ijrmhm.2012.02.021.
[54] H. Ledbetter, A. Migliori, A general elastic-anisotropy measure, *Journal of Applied Physics*, 100 (2006) 063516. https://doi.org/10.1063/1.2338835.
[55] C.M. Kube, Elastic anisotropy of crystals, *AIP Advances*, 6 (2016) 095209. https://doi.org/10.1063/1.4962996.
[56] R. Gaillac, P. Pullumbi, F.-X. Coudert, ELATE: an open-source online application for analysis and visualization of elastic tensors, *Journal of Physics: Condensed Matter*, 28 (2016) 275201. https://doi.org/10.1088/0953-8984/28/27/275201.
[57] A. Marmier, Z.A.D. Lethbridge, R.I. Walton, C.W. Smith, S.C. Parker, K.E. Evans, ElAM: A computer program for the analysis and representation of anisotropic elastic properties, *Computer Physics Communications*, 181 (2010) 2102–2115. https://doi.org/10.1016/j.cpc.2010.08.033.
[58] Y. Li, The anisotropic behavior of Poisson's ratio, Young's modulus, and shear modulus in hexagonal materials, *physica status solidi (a)*, 38 (1976) 171–175. https://doi.org/10.1002/pssa.2210380119.
[59] V.A. Gorodtsov, D.S. Lisovenko, Extreme values of Young's modulus and Poisson's ratio of hexagonal crystals, *Mechanics of Materials*, 134 (2019) 1–8. https://doi.org/10.1016/j.mechmat.2019.03.017.
[60] T.C.T. Ting, T. Chen, Poisson's ratio for anisotropic elastic materials can have no bounds, *The Quarterly Journal of Mechanics and Applied Mathematics*, 58 (2005) 73–82. https://doi.org/10.1093/qjmamj/hbh021.
[61] G.N. Greaves, A.L. Greer, R.S. Lakes, T. Rouxel, Poisson's ratio and modern materials, *Nature Materials*, 10 (2011) 823–837. https://doi.org/10.1038/nmat3134.
[62] Jahid Hassan, M.A. Masum, Ruman Ali, Md. Enamul Haque, S.H. Naqib, DFT based comparative study of the physical properties of *M*AlB (*M* = V, Ta, Mo, Nb) MAB compounds, *Next Materials*, 13 (2026) 102893. 10.1016/j.nxmate.2026.102893.
[63] M.A. Ali, M.M. Hossain, A.K.M.A. Islam, S.H. Naqib, Ternary boride $Hf_3PB_4$: Insights into the physical properties of the hardest possible boride MAX phase, *Journal of Alloys and Compounds*, 857 (2021) 158264. https://doi.org/10.1016/j.jallcom.2020.158264.
[64] O.L. Anderson, A simplified method for calculating the Debye temperature from elastic constants, *Journal of Physics and Chemistry of Solids*, 24 (1963) 909–917. https://doi.org/10.1016/0022-3697(63)90067-2.
[65] M.E. Fine, L.D. Brown, H.L. Marcus, Elastic constants versus melting temperature in metals, *Scripta Metallurgica*, 18 (1984) 951–956. https://doi.org/10.1016/0036-9748(84)90267-9.

[66] D.R. Clarke, Materials selection guidelines for low thermal conductivity thermal barrier coatings, *Surface and Coatings Technology*, 163–164 (2003) 67–74. https://doi.org/10.1016/S0257-8972(02)00593-5.
[67] G.A. Slack, Nonmetallic crystals with high thermal conductivity, *Journal of Physics and Chemistry of Solids*, 34 (1973) 321–335. https://doi.org/10.1016/0022-3697(73)90092-9.
[68] V.N. Belomestnykh, E.P. Tesleva, Interrelation between anharmonicity and lateral strain in quasi-isotropic polycrystalline solids, *Technical Physics*, 49 (2004) 1098–1100. https://doi.org/10.1134/1.1787679.
[69] C.L. Julian, Theory of heat conduction in rare-gas crystals, *Physical Review*, 137 (1965) A128–A137. https://doi.org/10.1103/PhysRev.137.A128.
[70] T. Ito, D. Pinek, T. Fujita, M. Nakatake, S. -i. Ideta, K. Tanaka, T. Ouisse, Electronic structure of $Cr_2AlC$ as observed by angle-resolved photoemission spectroscopy, *Physical Review B*, 96 (2017) 195168. https://doi.org/10.1103/PhysRevB.96.195168.
[71] D. Pinek, T. Ito, M. Ikemoto, K. Yaji, M. Nakatake, S. Shin, T. Ouisse, Unified description of the electronic structure of $M_2AC$ nanolamellar carbides, *Physical Review B*, 100 (2019) 075144. https://doi.org/10.1103/PhysRevB.100.075144.
[72] D. Pinek, T. Ito, K. Furuta, A.J. van Bunningen, P. Le Fèvre, F. Bertran, T. Ouisse, Fermi surface and band structure of $Ti_2SnC$ as observed by angle-resolved photoemission spectroscopy, *Physical Review B*, 104 (2021) 195118. https://doi.org/10.1103/PhysRevB.104.195118.
[73] I.R. Shein, A.L. Ivanovskii, Structural, elastic, electronic properties and Fermi surface for superconducting $Mo_2GaC$ in comparison with $V_2GaC$ and $Nb_2GaC$ from first principles, *Physica C: Superconductivity*, 470 (2010) 533–537. https://doi.org/10.1016/j.physc.2010.04.010.
[74] Y.M. Blanter, M.I. Kaganov, A.V. Pantsulaya, A.A. Varlamov, The theory of electronic topological transitions, *Physics Reports*, 245 (1994) 159–257. https://doi.org/10.1016/0370-1573(94)90103-1.
[75] M. Magnuson, M. Mattesini, Chemical bonding and electronic-structure in MAX phases as viewed by X-ray spectroscopy and density functional theory, *Thin Solid Films*, 621 (2017) 108–130. https://doi.org/10.1016/j.tsf.2016.11.005.
[76] Y. Bai, X. Qi, A.I. Duff, N. Li, F. Kong, X. He, R. Wang, W.E. Lee, Density functional theory insights into ternary layered boride MoAlB, *Acta Materialia*, 132 (2017) 69–81. https://doi.org/10.1016/j.actamat.2017.04.031.
[77] N. Ni, H. Zhang, Y. Zhou, Electronic structure, bonding characteristics, and mechanical behaviors of a new family of Si-containing damage-tolerant MAB phases $M_5SiB_2$ (M = IVB–VIB transition metals), *Journal of Advanced Ceramics*, 11 (2022) 1626–1640. https://doi.org/10.1007/s40145-022-0636-9.
[78] H. Yin, X. He, J. Zhang, G. Song, Y. Zheng, Y. Bai, DFT-assisting discovery and characterization of a hexagonal MAB-phase $V_3PB_4$, *Journal of Materiomics*, 9 (2023) 1141–1150. https://doi.org/10.1016/j.jmat.2023.07.002.
[79] F.-Z. Dai, Z. Feng, Y. Zhou, First-principles investigation on the chemical bonding, elastic properties and ideal strengths of MoAlB and WAlB nanolaminated MAB phases, *Computational Materials Science*, 147 (2018) 331–337. https://doi.org/10.1016/j.commatsci.2018.02.033.
[80] Y. Mo, P. Rulis, W.Y. Ching, Electronic structure and optical conductivities of 20 MAX-phase compounds, *Physical Review B*, 86 (2012) 165122. https://doi.org/10.1103/PhysRevB.86.165122.
[81] G. Hug, P. Eklund, A. Orchowski, Orientation dependence of electron energy loss spectra and dielectric functions of $Ti_3SiC_2$ and $Ti_3AlC_2$, *Ultramicroscopy*, 110 (2010) 1054–1058. https://doi.org/10.1016/j.ultramic.2010.05.007.

[82] S. Li, R. Ahuja, M.W. Barsoum, P. Jena, B. Johansson, Optical properties of $Ti_3SiC_2$ and $Ti_4AlN_3$, *Applied Physics Letters*, 92 (2008) 221907. https://doi.org/10.1063/1.2938862.
[83] Mst. Bina Aktar, F. Parvin, A.K.M. Azharul Islam, S.H. Naqib, Structural, elastic, electronic, bonding, thermo-mechanical and optical properties of predicted NbAlB MAB phase in comparison to MoAlB: DFT based ab-initio insights, *Results in Physics*, 52 (2023) 106921. https://doi.org/10.1016/j.rinp.2023.106921.
[84] M.I. Naher, S.H. Naqib, A comprehensive study of the thermophysical and optoelectronic properties of $Nb_2P_5$ via *ab-initio* technique, *Results in Physics*, 28 (2021) 104623. https://doi.org/10.1016/j.rinp.2021.104623.
[85] A. Togo, I. Tanaka, First principles phonon calculations in materials science, *Scripta Materialia*, 108 (2015) 1–5. https://doi.org/10.1016/j.scriptamat.2015.07.021.
[86] D. Alfè, PHON: A program to calculate phonons using the small displacement method, *Computer Physics Communications*, 180 (2009) 2622–2633. https://doi.org/10.1016/j.cpc.2009.03.010.
[87] I. Pallikara, P. Kayastha, J.M. Skelton, L.D. Whalley, The physical significance of imaginary phonon modes in crystals, *Electronic Structure*, 4 (2022) 033002. https://doi.org/10.1088/2516-1075/ac78b3.
[88] X. Gonze, C. Lee, Dynamical matrices, Born effective charges, dielectric permittivity tensors, and interatomic force constants from density-functional perturbation theory, *Physical Review B*, 55 (1997) 10355–10368. https://doi.org/10.1103/PhysRevB.55.10355.
[89] A. Champagne, F. Bourdarot, P. Bourges, P. Piekarz, D. Pinek, I. Gélard, J.-C. Charlier, T. Ouisse, Phonon dispersion curves in $Cr_2AlC$ single crystals, *Materials Research Letters*, 6 (2018) 378–383. https://doi.org/10.1080/21663831.2018.1463298.
[90] A. Togo, L. Chaput, I. Tanaka, G. Hug, First-principles phonon calculations of thermal expansion in $Ti_3SiC_2$, $Ti_3AlC_2$, and $Ti_3GeC_2$, *Physical Review B*, 81 (2010) 174301. https://doi.org/10.1103/PhysRevB.81.174301.
[91] Y. Bai, X. He, R. Wang, Lattice dynamics of Al-containing MAX-phase carbides: a first-principle study, *Journal of Raman Spectroscopy*, 46 (2015) 784–794. https://doi.org/10.1002/jrs.4720.
[92] G. Surucu, B. Yildiz, A. Erkisi, X. Wang, O. Surucu, The investigation of electronic, anisotropic elastic and lattice dynamical properties of MAB phase nanolaminated ternary borides: $M_2AlB_2$ (M = Mn, Fe and Co) under spin effects, *Journal of Alloys and Compounds*, 838 (2020) 155436. https://doi.org/10.1016/j.jallcom.2020.155436.